\documentclass[%
 reprint,
superscriptaddress,
nofootinbib,
 amsmath,amssymb,
 aps,
]{revtex4-2}

\usepackage[colorlinks,citecolor=blue]{hyperref}
\usepackage{graphicx}
\usepackage{dcolumn}
\usepackage{bm}
\usepackage[dvipsnames]{xcolor}
\usepackage{placeins}
\usepackage{soul}
\usepackage{lineno}
\definecolor{red}{rgb}{0.9, 0,0}
\definecolor{cerulean}{rgb}{0., 0.62,0.9}
\definecolor{navy}{rgb}{0.05, 0.05,0.8}

\newcommand{\heta}{\tilde\eta}
\newcommand{\hsigma}{\tilde\sigma}
\newcommand{\homega}{\tilde\omega}

\newcommand{\hpi}{\tilde\pi}
\newcommand{\hLambda}{\tilde\Lambda}
\newcolumntype{C}[1]{>{\centering\let\newline\\\arraybackslash\hspace{0pt}}m{#1}}

\begin{document}
\title{Perturbative benchmark models for a dark shower search program}

\author{Simon Knapen}
\affiliation{CERN, Theoretical Physics Department, Geneva, Switzerland}
\author{Jessie Shelton}
\affiliation{Illinois Center for Advanced Studies of the Universe, Department of Physics, University of Illinois, Urbana, IL 61801}
\author{Dong Xu}
\affiliation{Illinois Center for Advanced Studies of the Universe, Department of Physics, University of Illinois, Urbana, IL 61801}

\begin{abstract}
We provide five benchmark hidden valley models with perturbative parton showers, which span a wide range of dark shower phenomenology. We consider production through an $s$-channel, heavy mediator, which can be identified with the SM Higgs. By assuming a set of well-motivated decay portals, one can moreover fix both the branching ratios of the dark mesons and set a lower bound on their lifetime. We provide a public python tool which can be used to generate self-consistent PYTHIA 8 cards for our benchmark models.
\end{abstract}

\maketitle
\tableofcontents

\section{Introduction}
\label{sec:intro}

The possible collider signatures of confining dark sectors were first considered in the context of hidden valley models \cite{Strassler:2006im}, where it was noted that the resulting spectacular phenomenologies could prove to be well-motivated ``stress tests'' for the existing analysis strategies at the LHC.   A growing body of work has further developed hidden valleys as ingredients in models that address long-standing mysteries of particle physics, such as the stability of the electroweak hierarchy \cite{Chacko:2005pe, Burdman:2006tz,Craig:2015pha}, the matter-antimatter asymmetry \cite{Bai:2013xga}, dark matter \cite{Hur:2007uz,Kribs:2009fy,Beauchesne:2018myj,Francis:2018xjd,Bernreuther:2019pfb}, and the origin of neutrino masses \cite{Grossman:2010iq,Chacko:2020zze}.   More generally, the collider phenomenology of light, strongly-coupled hidden sectors is extremely rich and the subject of a  substantial amount of work, which includes studies focusing on displaced signatures \cite{Schwaller:2015gea,Curtin:2015fna,Csaki:2015fba,Pierce:2017taw,Alipour-Fard:2018lsf,Alipour-fard:2018mre,Bernreuther:2020xus,Yuan:2020eeu}, event or jet shape variables \cite{Harnik:2008ax,Cohen:2015toa,Knapen:2016hky,Cohen:2017pzm,Beauchesne:2017yhh,Park:2017rfb,Cohen:2020afv,Cesarotti:2020hwb,Cesarotti:2020uod,Cesarotti:2020ngq} and machine learning techniques \cite{Heimel:2018mkt,Bernreuther:2020vhm}, as well as more general explorations and model-building efforts \cite{Han:2007ae,Strassler:2006qa,Strassler:2008bv,Strassler:2008fv,Baumgart:2009tn,Luo:2009kf,Juknevich:2009ji,Juknevich:2009gg,Bunk:2010gb,Carloni:2010tw,Carloni:2011kk,Hur:2011sv,Cheng:2015buv,Englert:2016knz,Cheng:2016uqk,Lichtenstein:2018kno,Cheng:2018gvu,Renner:2018fhh,Cheng:2019yai,Alimena:2019zri,Beauchesne:2019ato}.

One of the more interesting and challenging features of hidden valley models is that they will generically produce an appreciable and variable multiplicity of dark particles, which are not necessarily isolated from each other.  Such high multiplicity, non-isolated signals have become known as ``dark shower'' topologies.  Dark showers pose special challenges in both prompt and displaced scenarios.  In the prompt case the main challenge is to extract dark shower signatures from large QCD backgrounds.  Background rejection can often be easier in the displaced scenario, where at least one particle has a detector-scale lifetime, but in this case experimental analyses are often challenging, while theoretical uncertainties become substantially more confounding to address.

Though the majority of experimental searches for displaced objects so far have focused on events with one or two isolated displaced objects, CMS has also performed a dedicated analysis for the ``emerging jet'' topology \cite{Sirunyan:2018njd}, which is a dark shower topology with a large number of overlapping displaced decays before the calorimeters.  In developing a search strategy for dark showers, some important overarching goals are to: (i) assess the extent to which existing searches for low-multiplicity, isolated displaced objects are sensitive to this class of models; 
(ii) evaluate the robustness of searches such as \cite{Sirunyan:2018njd} under the variation of the theoretical assumptions used to generate the signal models; and (iii) identify new experimental strategies that can expand coverage of the signature space. In this paper we hope to take a step toward these goals by investigating the theoretical conditions under which a dark shower topology is realized in a set of well-motivated benchmark models. In particular, to obtain a dark shower, the lifetime of the dark sector mesons should not parametrically exceed the size of the detector, as in this case the discovery signature is usually a single \emph{isolated} displaced vertex. A growing number of existing and planned searches for displaced objects has sensitivity to the latter scenario \cite{Alimena:2019zri}, which is not the primary subject of this paper.

There now exist a good number of interesting examples of hidden valley models that produce a variety of dark shower topologies, as referenced above. These various examples may however be singular points in the viable model space, and it is not yet clear how representative their phenomenology is in the vast space of possibilities. The reasons for this are:
\begin{enumerate}

\item One often finds a quite delicate model dependence in the properties of the new states being produced and/or detected. Unlike e.g.~supersymmetry, there also is no well-defined, finite set of model parameters one can explore systematically.

\item It is difficult and often impossible to predict the detailed dynamics in the hidden sector, in particular in the context of incalculable hadronization processes.

\item The experimental challenges in separating signals from backgrounds are severe and subtle, as signals are often soft and/or not isolated. This in particular poses trigger challenges, which inevitably bias search strategies and therefore the classes of models that one searches for. It also means that experimental searches for dark showers tend to be complicated and labor-intensive.

\end{enumerate}

Point (iii) in particular calls for an unbiased ``simplified models'' strategy, where toy models are constructed that capture the main phenomenological features and can be recast most easily, with minimal reference to theoretical priors. This approach was applied with great success in the context of supersymmetry \cite{Alves:2011wf} and dark matter searches \cite{Abdallah:2015ter}. Points (i) and (ii) however imply that feasibility of such a program in the context of dark shower topologies is doubtful. As a step towards a more comprehensive experimental coverage, we therefore follow a hybrid approach: we present a fairly small set of simplified models which are meant to capture a broad range of \emph{phenomenological} features, but we do inject some minimal theory priors in order to arrive at more concrete and predictive benchmarks, with a smaller number of free parameters. We supply a small python package available at 
\begin{center}
{\small
\url{https://gitlab.com/simonknapen/dark_showers_tool}}
\end{center}
that encodes the theory priors described in the remainder of this paper and can be used to generate self-consistent configuration files for the PYTHIA 8 hidden valley module \cite{Sjostrand:2014zea,Carloni:2010tw,Carloni:2011kk}. 

We carefully explain and justify our theory priors in Sec.~\ref{sec:rightopinions}. While we believe that our set of benchmarks could be a valuable next step in this fledging experimental program, we also clearly highlight its obvious limitations: out of practical necessity, the present work focuses on models with \emph{perturbative} dark showers, as our level of theoretical control for non-perturbative dark showers is significantly more limited. We also use a highly simplified hadronization model, which does not capture potential cascade decays among the dark sector mesons. We define and discuss our benchmark models in Sec.~\ref{sec:benchmarks} and conclude in Sec.~\ref{sec:discussion}. Details regarding the treatment of hadronization and the UV completions of the various decay portals are left for the appendices.

\section{Theory priors}
\label{sec:rightopinions}

Following \cite{Alimena:2019zri}, we can decompose a typical dark shower topology into three parts:
\begin{enumerate}
\item Production: A heavy state ($\gg$ GeV) is produced that subsequently decays to lighter states in the dark sector.  This state may be a known particle, such as the SM Higgs, or a new BSM mediator.

\item Evolution within the dark sector, i.e.~the shower itself: Strong dynamics give rise to a dark shower, yielding a final dark hadron state with variable and often large particle multiplicity, and a non-trivial phase space distribution. 

\item Dark hadron decay: Dark hadrons may decay back to the SM with a wide variety of possible lifetimes and final states.
\end{enumerate}
Both production and decay steps involve interactions with the SM, where it is possible to make some definite statements about the allowed possibilities. 
Evolution within the dark sector, however, is challenging to describe in full quantitative detail.

In constructing benchmark models for experimental searches, the availability of Monte Carlo simulation tools is a major consideration.  We will accordingly focus our attention on theories that, like QCD, admit a perturbative parton shower description, as implemented in the PYTHIA 8 hidden valley module \cite{Carloni:2010tw,Carloni:2011kk}. 

\subsection{Production}   

A dark shower generically begins with the pair production of particles charged under the dark gauge group.  There are a wide variety of possibilities for this initial production step, itemized in \cite{Alimena:2019zri}.  These possibilities may be divided into two categories: 
\begin{itemize}
\item  resonant production of SM-singlet dark partons $\psi$ through the decay of a massive mediator, $X\to \bar \psi \psi$.   Attractive choices for the massive mediator are a new vector boson
 \cite{Strassler:2006im,Han:2007ae,Strassler:2008bv,Strassler:2008fv,Baumgart:2009tn,Cohen:2015toa,Beauchesne:2018myj,Bernreuther:2019pfb,Pierce:2017taw,Bernreuther:2020xus,Cohen:2017pzm,Park:2017rfb,Yuan:2020eeu}, 
 a new scalar boson 
 \cite{Strassler:2008bv,Hur:2011sv,Knapen:2016hky,Englert:2016knz,Beauchesne:2017yhh,Beauchesne:2018myj,Alipour-fard:2018mre},  
 the SM Higgs 
\cite{Strassler:2006im,Strassler:2006ri,Hur:2007uz,Strassler:2008bv,Juknevich:2009gg,Craig:2015pha,Csaki:2015fba,Craig:2016kue,Curtin:2015fna,Alipour-Fard:2018lsf} or the SM $W/Z$ bosons  \cite{Strassler:2006im,Cheng:2019yai,Chacko:2020zze}.  In this class of models, the mediator mass sets the overall energy scale of the event. Stringent constraints on dilepton (and secondarily dijet) resonances require a new vector mediator to have TeV-scale masses if its branching ratios to SM particles and dark partons are comparable, but much lighter masses remain viable when SM branching ratios are suppressed.

\item non-resonant pair production of a new particle $X_D$ that is charged under both SM and dark gauge groups.  Appreciable production cross-sections can readily be obtained when $X_D$ carries color or electroweak charge, as in \cite{Strassler:2006qa,Carloni:2010tw,Carloni:2011kk,Cheng:2015buv,Schwaller:2015gea,Cheng:2016uqk,Lichtenstein:2018kno,Cheng:2018gvu,Renner:2018fhh}. Especially for colored mediators, LHC constraints then require $X_D$ to be TeV-scale.  This production portal therefore predicts energetic events, with a number of hard SM jets and/or a large amount of $H_T$.  
\end{itemize}
The choice of production mode determines the overall mass scale of the event, the types of accompanying SM objects, and the expected event rate.

 From the point of view of designing an inclusive and systematic search program for dark showers, production through the SM Higgs  stands out as a particularly simple and compelling scenario.  
 The SM Higgs is one of our most sensitive windows onto low-mass, SM singlet states \cite{Curtin:2013fra}, making Higgs-initiated decays especially well motivated; indeed, Higgs decays into dark showers are the leading discovery signal of a class of solutions to the hierarchy problem \cite{Craig:2015pha,Curtin:2015fna}.
 Taking the SM Higgs boson to be the initiating mediator has the great advantage of using a particle that is already known and characterized.  The Higgs mass is fixed and its production cross-sections are well-understood.  Its branching fraction into dark showers is bounded by $\text{Br}(h \to\mathrm{exotic})< 0.21$ at 95\% CL \cite{Aad:2019mbh}, but any smaller value can be a consistent possibility. 

As  production through the SM Higgs results in relatively low-mass final states compared to the LHC center-of-mass energy, this choice is a challenging but especially well-motivated option.  Searches that are capable of discovering Higgs-initiated dark showers can likely be repurposed to also be sensitive to models initiated by much heavier mediators; the reverse is far from guaranteed. Thus using SM Higgs production to guide the development of a generic dark shower search strategy has the great advantage of building a comprehensive net capable of catching dark showers produced in a wide range of scenarios. 

Because high-multiplicity exotic Higgs decays are notoriously challenging signals, the Higgs-initiated scenario may ultimately prove too ambitious in some cases. We therefore also consider production through the decay of a new, heavier $s$-channel mediator, which injects more energy into the dark sector than the SM Higgs and thus generates a more striking signature. Some of the decay portals we consider moreover \emph{require} the existence of some such additional state and in these cases we will assume that the decay of these new particles initiates the dark shower, as explained in Sec.~\ref{sec:benchmarks}. To lighten the notation, we will use the uniform notation $H$ for the heavy particle that is responsible for initiating the dark shower, with the understanding that whenever we set \mbox{$m_H=125$ GeV}, we identify $H$ with the SM Higgs boson.  We focus on $s$-channel production mechanisms since, unlike  $t$-channel and related production mechanisms that rely on the SM charges of BSM particles, $s$-channel production does not predict accompanying SM objects in the final state.  Searches for the benchmark models  we provide here must therefore rely centrally on the objects produced in the dark shower itself.  These benchmark models are thus more challenging but more generic targets for the development of a more inclusive dark shower search program.

\subsection{Showering and hadronization\label{sec:showerandhadron}}   

 A key feature of dark shower signatures is the nontrivial evolution of energy within the dark sector that follows the initial production at energy scale $Q$. In QCD, the patterns of energy flow and the associated {\em parton} multiplicities are dominated by the soft and collinear singularities of the theory, and can be described using perturbation theory.  This feature holds generally for theories that, like QCD, have small 't Hooft couplings $\lambda \equiv g^2 N_c$, with $g$ and $N_c$ the gauge coupling and number of colors, at the scale $Q$ at which the shower is initiated. In these theories $\lambda$ can become large, but only in a limited energy range near the confinement scale $\Lambda$.  Thus in QCD non-perturbative contributions to the overall energy flow in an event are suppressed by the small quantity $\Lambda/Q$: e.g., non-perturbative contributions to a jet's $p_T$ are of order  $\Lambda$ (see, e.g., \cite{Salam:2009jx} for a review).   In the small-$\lambda$, ``QCD-like'' regime, well-established parton shower algorithms allow for good modeling of the partonic component of the hidden sector evolution.   
 
 While parton showers are under reasonably good control at small $\lambda$, the process of hadronization is not. 
 To fully describe the final state produced by a dark shower, we need to know the dark hadron spectrum as well as the multiplicities of those hadrons that are produced in a given event.  
 The full mass spectrum of a confining hidden sector is in general poorly understood, with detailed results available (e.g. from lattice calculations) for only a few specific theories \cite{Morningstar:1999rf,Francis:2018xjd,DeGrand:2019vbx,Kribs:2016cew}.  
 But even given a spectrum, reliably computing the multiplicities of different dark hadrons produced in an event is beyond the reach of our current tools:  The total multiplicity of a specific hadron species is not an infrared-collinear-safe observable, meaning it is a quantity that is inherently sensitive to the nonperturbative dynamics of the hidden sector.  For small-$\lambda$ hidden sectors where all hadron decays proceed promptly, searches for dark showers can circumvent this issue by focusing on IRC-safe observables for which theory predictions are under perturbative control \cite{Cohen:2020afv}.  However, as soon as one or more species develop detector-scale lifetimes, as is the case for example for both emerging and  semi-visible jet topologies, the detector signature becomes inextricably intertwined with incalculable properties of the model.   For semi-visible jets, these incalculable hadronic uncertainties can be absorbed into a single parameter describing the fraction of invisible particles  \cite{Cohen:2015toa}.  In signatures where the visible multiplicity of a specific species becomes of primary importance, as is the case in searches for displaced final states, the incalculable nature of hadron multiplicity must be confronted head-on. In other words, in such cases the model is effectively \emph{defined} in part by the assumptions made about hadronization, and for results to be reproducible it is necessary to be very explicit about the settings used to generate the signal Monte Carlo. Monte Carlo tools moreover evolve in time, which may make it difficult to exactly reproduce the event generation many years after the original analysis. We therefore recommend that experimental collaborations provide the most relevant truth-level distributions, such as multiplicity, $p_T$ spectra, etc., in their supplementary material.
 
In theories with $\lambda \gg 1$,  soft and collinear splittings are no longer parametrically preferred over more general, wide-angle emission.  Insight into the behavior of gauge theories at very large $\lambda$ can be obtained using the AdS-CFT correspondence, and extends to some confining theories \cite{Polchinski:2002jw, Klebanov:2000hb}.   In particular, theories where $\lambda$ remains large over an extended range of energies are expected to yield spherical event shapes \cite{Strassler:2008bv, Csaki:2008dt}, a result which has been directly proven for a class of conformal theories at large $\lambda$ \cite{Lin:2007fa,Hofman:2008ar,Hatta:2008qx}. Similar ideas have been explored in the context of Randall-Sundrum constructions \cite{Costantino:2020msc}.  When at least some of the final (massive) particles in such theories can decay into the SM, such spherical events yield ``soft bombs'', or ``soft unclustered energy patterns (SUEPs)'' \cite{Strassler:2008bv,Knapen:2016hky}. Only for this specific and idealized scenario is a Monte Carlo tool currently available \cite{suepcode}. 

In the intermediate regime where $\lambda\sim 1$ over a substantial portion of the dark sector evolution, there are no known theoretical handles.  Work by Cesarotti, Freytsis, Reece and Strassler in Refs.~\cite{Alimena:2019zri,Cesarotti:2020uod,Cesarotti:2020ngq} has supplied a phenomenological interpolation between perturbative parton showers and showers based on gauge-gravity duals in the $\lambda\sim 1$ regime, where both techniques are (deliberately) pushed out of their regime of validity. The predictions in this regime are therefore to be understood as toy models that interpolate between two regimes in which there is theoretical control.  With those caveats, the analysis in \cite{Alimena:2019zri} found that predictions for nonperturbative quantities such as the hadron multiplicity can be notably different in different toy models. The ideal search strategy would therefore be maximally inclusive and would strive to avoid making unnecessary assumptions about the distribution of objects, either spatially or in energy. How to best implement these principles given the existing Monte Carlo tools and trigger limitations is currently still an open question, and requires a continued dialogue between the experimental and theory communities.

To develop our benchmark models, we will stay within the relatively well-understood regime of QCD-like theories, where we can take full advantage of existing resources for event generation.   
In particular, our benchmarks are constructed using the Hidden Valley module \cite{Carloni:2010tw,Carloni:2011kk}
within PYTHIA 8 \cite{Sjostrand:2014zea}.  The Hidden Valley module describes showers and hadronization in a $SU(N_c)$ gauge theory with $N_\psi$ fundamentals, which may be taken to be scalars or fermions. Our benchmarks, further detailed in the next section, fix $N_c=3$ and $N_\psi=1$.
While we endeavor to make our benchmark models as theoretically sensible as possible, we emphasize that they should be considered as useful signature generators that are broadly illustrative of a general class of models, rather than completely internally consistent theories.  
Even within the parton shower component of the evolution, the PYTHIA Hidden Valley shower is strictly a leading-log algorithm, and the missing higher-order corrections are a notable source of systematic uncertainty  \cite{Cohen:2020afv}.

Even in QCD-like theories, modeling hidden hadronization introduces additional and less quantifiable uncertainties. 
In the PYTHIA Hidden Valley module, hadronization proceeds via a Lund string model \cite{Andersson:1983ia} to a highly simplified hadron sector. This stripped-down hadron sector may or may not be reflective of the low-lying hadron dynamics of a given theory.  Concretely, the PYTHIA hidden valley model includes only spin-zero and spin-one mesons, which can each be diagonal or off-diagonal with respect to a dark flavor symmetry (the analogue of isospin); baryons and all excited mesons are absent. The SM $\pi^0$, $\pi^\pm$, $\rho^0$ and $\rho^\pm$ are the closest analogues to the states provided in the hidden valley module, though this analogy does not necessarily hold for the permissible mass spectrum. In particular the (breaking of the) flavor symmetries in the dark sector may be very different from the SM flavor structure. Moreover, the parameters of the Lund string model, which control hadron multiplicities, may be tuned to data in the case of QCD, but this is hardly possible in a hidden valley model.   Therefore both the overall hadron multiplicity and the relative multiplicities of different individual hadron species within our benchmark models are highly uncertain, and the predictions from PYTHIA should not be taken literally.  The authors of the PYTHIA hidden valley module are fully aware of these uncertainties, and offer the user a chance to set by hand both the mass hierarchy  between spin-zero and spin-one mesons as well as the relative fractions of spin-zero and spin-one species produced in hadronization.  

Given the limitations outlined above, our benchmarks will necessarily be toy models, but we will attempt to use the freedom given to us by PYTHIA to construct toys that approximate the key features of more realistic scenarios. Concretely, we will set $N_\psi=1$, which implies that the PYTHIA hadron spectrum consists of a spin-zero and a spin-one meson, which do not carry any flavor symmetry.\footnote{In our python package we however allow the user to specify $N_\psi>1$, in which case all spin-zero, off-diagonal dark mesons, the analogues of the $\pi^\pm$ in the SM, are taken to be detector-stable. They can be made to decay by modifying the PYTHIA card by hand, but this feature is  not currently implemented in the python package. We emphasize that this too is a toy model, and does not accurately reflect most $N_\psi>1$ theories. } We denote these particles with $\heta$ and $\homega$ respectively, in loose analogy with their SM counterparts. If the $\homega\to\heta\heta$ channel is kinematically open, we assume that this decay occurs promptly with 100\% branching ratio. These choices capture salient features of the phenomenology of a  $N_\psi=1$ gauge theory, as well as theories with $N_\psi>1$ for which the flavor symmetries are maximally broken. In Appendix~\ref{sec:pythia} we elaborate on  which features of realistic theories this approach does and does not capture.

As this highly simplified spectrum is merely a stand-in for the more complex spectra predicted in full theories, we treat $m_{\heta}$, $m_{\homega}$ and  confinement scale ($\hLambda$) as independent parameters. In the python package they can be accessed by setting the parameters $m_{\heta}$, $\xi_\omega\equiv m_{\homega}/m_{\heta}$ and $\xi_\Lambda\equiv\hLambda/m_{\heta}$ for all models, except the vector portal model, for which $\xi_\Lambda\equiv\hLambda/m_{\homega}$. Our default choices are $\xi_\Lambda=\xi_\omega=1$, unless stated otherwise.  Since the $\homega$ has higher spin than the $\heta$, we expect $m_{\homega}\approx \hLambda \gtrsim m_{\heta}$, with $m_{\heta}\to 0$ in the $N_c\to\infty$ limit \cite{Witten:1979vv}. Choices that deviate strongly from this expectation could be considered as theoretically less probable. Our notation is summarized in Tab.~\ref{tab:notation}.
\begin{table}
\begin{tabular}{c|l}
$\hLambda$& dark sector confinement scale\\
$\heta$& dark sector spin-zero meson\\
$\homega$& dark sector spin-one meson\\
$\psi$ & dark sector quark\\
$\xi_\omega$& $m_{\homega}/m_{\heta}$\\
$\xi_\Lambda$& $\hLambda/m_{\homega}$ for the vector portal and  $\hLambda/m_{\heta}$ otherwise\\
$N_c$& number of dark sector colors \\
$N_\psi$&number of dark sector flavors \\
$a$ & elementary pseudoscalar mediator (Secs.~\ref{sec:gluon_alp_portal} and \ref{sec:photon_alp_portal})\\
$A'$ & elementary vector mediator (Secs.~\ref{sec:vector_portal} and \ref{sec:lightDP})
\end{tabular}
\caption{Summary of notation used in this paper and the in the python package. The additional elementary particles $a$ and $A'$ are required in some models, and are introduced below.\label{tab:notation}}
\end{table}
We  set the relative multiplicity of $\heta$ and $\homega$ mesons according to a toy hadronization model, which is based on QCD-informed expectations as described in Appendix~\ref{sec:pythia}. We address the hadronization uncertainty by selecting two benchmark values for the mass ratio $\xi_{\homega}$ where the $\homega\to 2\heta$ decay channel is respectively open and closed.  This choice has the effect of toggling between two different scenarios for the total multiplicity of dark mesons produced in a typical event.

In our benchmarks, the typical meson multiplicities produced by the PYTHIA hidden valley module are smaller than the typical multiplicities in an analogous QCD event. This is largely because we focus on the regime with $m_{\heta,\homega} \gtrsim 1$ GeV, primarily motivated by the lifetime considerations in the next section. The absence of complex cascade decays also contributes to the reduced multiplicity as compared to SM jets.

\subsection{Decay}

Depending on the symmetries of the dark sector, some hidden hadrons may be stable (making them dark matter candidates) while others will be able to decay through various portal interactions back to the SM.   Dark hadron decays in the regime of primary interest to us occur at low energies ($\lesssim$ 30 GeV),  where possible interactions with the SM are subject to numerous constraints from both energy and intensity frontier experiments.

\def\arraystretch{1.5}
\begin{table*}[!bth]\centering
\begin{tabular}{p{3.5cm}p{3.5cm}p{1.cm}p{3.cm}p{4.cm}p{1.0cm}}
Decay portal&decay operator & VDP&other dark hadron  & features& section\\\hline\hline
A. gluon portal&$\heta G^{\mu\nu}\tilde G_{\mu\nu}$ &$\heta$&$\homega$ stable or $\homega\to\heta\heta$ & hadron-rich shower&\ref{sec:gluon_alp_portal}\\
B. photon portal&$\heta F^{\mu\nu}\tilde F_{\mu\nu}$  &$\heta$&$\homega$ stable or $\homega\to\heta\heta$& photon shower&\ref{sec:photon_alp_portal}\\
C. vector portal& $\homega^{\mu\nu}F_{\mu\nu}$ &$\homega$&$\heta$ stable& semi-visible jet&\ref{sec:vector_portal}\\
D. Higgs portal&$\heta H^\dagger H$  &$\heta$&$\homega$ stable or $\homega\to\heta\heta$& heavy flavor-rich shower&\ref{sec:Higssmixing}\\
E. dark photon portal&$\heta F'^{\mu\nu}\tilde F'_{\mu\nu} + \epsilon F'^{\mu\nu} F_{\mu\nu}$  &$A'$&$\homega$ stable or $\homega\to\heta\heta$ &  lepton-rich shower& \ref{sec:lightDP}\\
\end{tabular}
\caption{Overview of the decay portals and their corresponding phenomenology. The $\heta$ and $\homega$ represent respectively the lightest spin-zero and spin-one meson in the dark sector. The $\heta$ is assumed to be a scalar in the Higgs portal model and a pseudoscalar in the remaining models. We further defined $\tilde F_{\mu\nu}\equiv \frac{1}{2}\epsilon_{\mu\nu\rho\sigma} F^{\rho\sigma} $ and $\tilde F'_{\mu\nu}\equiv \frac{1}{2}\epsilon_{\mu\nu\rho\sigma} F'^{\rho\sigma} $as the dual field strengths of the SM and dark photon respectively. The decay portal column indicates the operator(s) that allow the unstable dark meson to decay. The ``other dark hadron'' column indicates the possible decay channels for the hadron that does \emph{not} decay back to the SM. \label{tab:portals}}
\end{table*}

In this paper, we will assume that only a single species of dark hadron ($\heta$ or $\homega$) has a detector-relevant lifetime.  We will refer to the unstable particle as the \emph{visibly-decaying dark particle} (VDP).  The VDP is often, but not always, the lightest hadron in the dark sector.  In our simplified setup, other dark hadrons are assumed to decay promptly to the unstable hadron or to escape the detector as missing energy.  This choice is made for practical reasons, but (i) it is broadly representative of the phenomenology of a wide class of models, and (ii) it allows a wide range of qualitatively distinct signatures to be realized with a minimal number of arbitrary parameters. It is however manifestly insufficient to cover the \emph{full} range of possible dark shower topologies, as models with multiple species with different lifetimes are also well-motivated. The phenomenology in such showers could moreover be interesting experimentally, as the longest-lived species may provide a trigger in the muon system, where the shorter lived particles in the same event could result in multiple displaced vertices in the tracker. We leave such cases for a future study.

In describing the decays of the VDP, we invoke some simple theoretical considerations to winnow down the vast space of possibilities to a more restricted set of particular interest.  Concretely, we impose the following theory priors:
\begin{itemize}
\item No new sources of SM flavor violation.
\item In the infrared effective theory, we allow for operators up to dimension five. This set of operators includes commonly considered portals such as a dark meson mixing with the SM Higgs or photon, and the dimension-five coupling of a \mbox{(pseudo-)scalar} dark meson to SM photons or gluons.  Operators with higher mass dimension give rise to similar final states at the particle level, but (prohibitively) longer hadron lifetimes. 
\end{itemize}
These two criteria lead to the set of operators given by entries A to D in Tab.~\ref{tab:portals}, which represent three out of the four commonly considered ``portal'' operators that can connect a light dark sector with the SM (see, e.g.,\cite{Agrawal:2021dbo}).\footnote{ The fourth, the neutrino portal, is not considered here as it violates our first assumption. For hidden valley models with dark hadron decays to the SM via the neutrino portal, see \cite{Chacko:2020zze}.}. In addition, we consider one additional scenario that involves a new, light elementary degree of freedom:
\begin{itemize}
\item In Sec.~\ref{sec:lightDP} we consider a scenario which contains an elementary dark photon, in addition to the dark sector mesons (Entry E in Tab.~\ref{tab:portals}). The $\heta$ can then decay promptly via a dark chiral anomaly to a pair of dark photons, analogous to the $\pi^0$ in the SM.  The elementary dark photon $A'$ subsequently decays to the SM and is therefore the VDP in this scenario. \footnote{This scenario satisfies the two previous criteria both for the operator governing the decays of the $\heta$ as well as those of the elementary dark photon.}
\end{itemize}
This model allows for a  qualitatively \emph{shorter} minimal VDP lifetime than the other portals in Tab.~\ref{tab:portals}, as well as a higher particle multiplicity. It also yields displaced dilepton vertices without the MET that accompanies them in model C.  On these grounds, adding this benchmark scenario  allows us to cover additional phenomenological signatures that are distinct from the remaining scenarios.  The dynamics of the dark photon portal have a clear and compelling analogy with the  SM $\pi^0\to\gamma\gamma$ process, and are a built-in feature of models which have a dark copy of the SM, such as twin Higgs models \cite{Chacko:2005pe}. Although the  IR effective theory  describing the dark photon portal is less minimal, insofar as it requires an additional light degree of freedom relative to the other models, it is worth emphasizing that most of the other portals also require additional degrees of freedom  above the confinement scale in the dark sector in order to generate the portal operator, as we discuss further in the Appendix.  


These considerations lead us to a set of five scenarios, as summarized in Tab.~\ref{tab:portals}. Within each scenario, the branching ratios to SM states are completely determined,
depending only on the mass and (for particles decaying directly into hadrons) CP properties of the unstable hadron.  These benchmark scenarios moreover produce a wide range of interesting final state objects, as indicated in the ``features'' column in Tab.~\ref{tab:portals}. 
In this sense, we believe they provide good candidate benchmarks to help expand the existing searches towards a more inclusive search program.

\begin{figure}[t]
\includegraphics[width=0.47\textwidth]{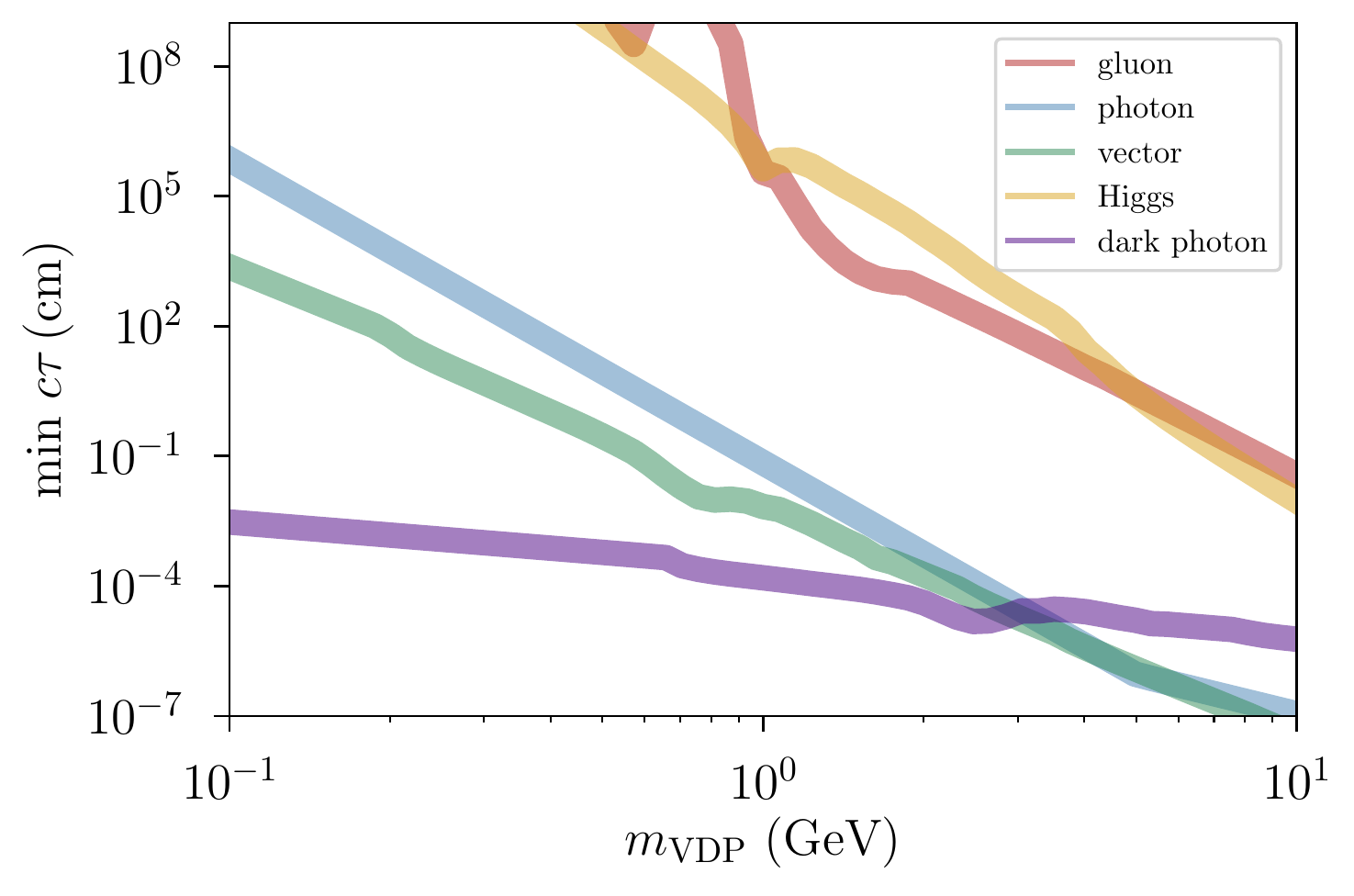}
\caption{Estimated minimum proper lifetime for the VDP, decaying through the various portals of Table~\ref{tab:portals}; see Sec.~\ref{sec:benchmarks} for details.
\label{fig:ctau}}
\end{figure}

In addition to fixing the branching ratios of the VDP, the choice of portal has important consequences for its expected lifetime.  
For the gluon and Higgs portal scenarios in particular, we find that {\em displaced decays} are generic in the $m_{\heta} \lesssim 15$ GeV regime.  Three factors contribute to make $\heta$ long-lived in these cases:
\begin{itemize}

\item Constraints on the portal couplings that appear in the IR effective theory from both energy and intensity frontier experiments.

\item Model-building considerations when generating the decay portal operator.  For instance, in order to generate a dominant, axion-like $\heta G\widetilde G$ coupling, new colored states are needed. The possible masses for such colored particles are however restricted by LHC constraints, which therefore indirectly limit the maximum strength of this operator.

\item Dark hadrons are composite states.  Thus (in our QCD-like benchmarks where anomalous dimensions are small) they are described by high-mass dimension operators in the ultraviolet theory, and their decay matrix elements are suppressed by the corresponding powers of $\hLambda/M$, where $M$ is a large mass scale associated with the UV completion. 
\end{itemize}
In Sec.~\ref{sec:benchmarks} and Appendix~\ref{sec:lifetime} we discuss these considerations in detail for the decay portals listed in Table~\ref{tab:portals}, deriving a \emph{lower bound} on the lifetime of the VDP in each scenario.   These lower bounds, shown in Fig.~\ref{fig:ctau}, are not rigorous exclusions, as they invoke theoretical biases in constructing relatively minimal UV completions, avoiding fine tuning. More elaborate UV completions or tuning could often achieve shorter lifetimes, though they render the model less plausible in our estimation. Rather, these lower bounds therefore indicate the kinds of lifetimes which can be thought of as most natural or generic.

In large-$\lambda$ theories one may well expect large anomalous dimensions for the operators that create dark hadrons, and in particular those that create spin-zero hadrons. In this case the dark hadron fields could appear more ``elementary'' than in the small-$\lambda$ regime, and their decays would accordingly be less suppressed.  However we emphasize that these shorter dark hadron lifetimes would be correlated with more isotropic average events, and that our choice to use PYTHIA hidden valley module means that we do not model this regime.

For the dark photon portal, the $\heta$ decays promptly to a pair of dark photons ($A'$), and the relevant lifetime constraints are on the (elementary) dark photon itself. Prompt dark photon decays are currently still allowed for $m_{A'}\gtrsim 10$ MeV \cite{Lees:2014xha,Aaij:2019bvg}, though planned searches at LHCb in particular \cite{Ilten:2015hya} would raise this minimum prompt mass to $m_{A'}\gtrsim 100$ MeV by the end of the high luminosity run of the LHC. 

Once $c\tau \gg 1$ m, the probability of observing an event with multiple decays in the detector becomes negligibly small. In such scenarios, the phenomenology of interest is that of a single displaced vertex, rather than of a dark shower.\footnote{If the mass of the VDP is low and $c\tau \gg 1\, m$, this scenario may be too soft for the main LHC detectors. On the other hand, due to the increased multiplicity of long-lived particles in these events, the sensitivity of possible external detectors such as MATHUSLA \cite{Chou:2016lxi} and CODEX-b \cite{Gligorov:2017nwh} is enhanced relative to the sensitivity for long-lived particles produced singly or in pairs \cite{Curtin:2018mvb}.} Given that our focus is on topologies with multiple dark sector meson decays occurring within the same event,  Fig.~\ref{fig:ctau} therefore implies an approximate lower bound on the mass of the VDP for which a dark shower signature is realized in each of the scenarios in Table~\ref{tab:portals}. These lower bounds are particularly relevant for the Higgs and gluon portal models and have important implications for the branching ratios of the $\heta$, which impact both online and offline search strategies.  For example, in the Higgs portal model we will see that the $\heta \to \tau\tau$ and $\heta\to c\bar c$ channels are always kinematically open for $m_{\heta}$ above our approximate lower bound.

\FloatBarrier

\section{Benchmark models\label{sec:benchmarks}}

\subsection{Gluon portal\label{sec:gluon_alp_portal}}

If the VDP has zero spin, it can be coupled to SM gluons through the operators
\begin{equation}\label{eq:gluonportal_bulk}
\mathcal{L}\supset\frac{\alpha_s}{8\pi}\frac{\heta}{f_{\heta}}  G_{\mu\nu}\tilde G^{\mu\nu}\quad\text{and/or}\quad\frac{\alpha_s}{8\pi}\frac{\heta}{f_{\heta}}\ G_{\mu\nu} G^{\mu\nu},
\end{equation}
depending on the CP properties of $\heta$. Here we defined $\tilde G^{\mu\nu}\equiv \frac{1}{2}\epsilon^{\mu\nu\rho\sigma}G_{\rho\sigma}$. The parameter $f_{\heta}$ is the effective decay constant of $\heta$ and sets its lifetime. We will work under the assumption that $\heta$ is a pseudo-scalar and thus couples through the first operator \eqref{eq:gluonportal_bulk} only. In CP-violating scenarios where both operators in (\ref{eq:gluonportal_bulk}) are present, the VDP lifetime can be altered by an $\mathcal{O}(1)$ amount as compared to the CP-conserving case.  This is particularly so for $m_{\heta}\lesssim 2$ GeV, where the various exclusive decay modes depend qualitatively on the CP properties of $\heta$.

The operator in \eqref{eq:gluonportal_bulk} is an irrelevant operator and therefore requires a UV completion, all the more so because $\heta$ itself is a composite particle.  As an intermediate step, we consider a theory with a heavy, elementary pseudo-scalar particle $a$ that mixes with the $\heta$-meson
\begin{equation}\label{eq:gluon_intermediate}
\mathcal{L}\supset-\frac{1}{2}m_a^2 a^2 -\frac{ \alpha_s}{8\pi}\frac{1}{f_a} a G_{\mu\nu}\tilde G^{\mu\nu}  - i y_\psi a \bar \psi \gamma_5 \psi 
\end{equation}
where $f_a$ and $y_\psi$ respectively parametrize the coupling of $a$ to the gluons and to the dark sector quarks $\psi$. This particular choice of mediator has the additional advantage that $a$ can also serve as an $s$-channel production portal, through the process $gg \to a \to \psi \bar\psi$, after which the $\psi$ initiate the shower in the dark sector.

With the normalization defined in \eqref{eq:gluonportal_bulk}, the {\em partonic} width of $\heta$ to gluon pairs at next-to-leading order is then given by \cite{Chetyrkin:1998mw}
\begin{equation}\label{eq:aglueglue}
\Gamma = \frac{\alpha_s(m_{\heta})^2}{32\pi^3}\frac{m_{\heta}^3}{f_{\heta}^2} \left[1+\left(\frac{97}{4}-\frac{7}{6}N_f\right)\frac{\alpha_s(m_{\heta})}{\pi}\right]
\end{equation} 
with $N_f$ the number of SM quarks with $m_f<m_{\heta}$. This formula is valid if $\heta$ is heavy enough that decays to gluons is a good approximation; following \cite{Aloni:2018vki}, we define this regime as $m_{\heta} >$ 1.84 GeV.  Here $\alpha_s(m_{\heta})$ is the running SM strong coupling, evaluated at $m_{\heta}$. For $m_{\heta}< 1.84$ GeV, we consider direct decays to hadrons following the treatment of Aloni et.al.~\cite{Aloni:2018vki}. We refer to Appendix~\ref{app:gluonportal} for more details and for its implementation in our python package.

In a full UV completion of \eqref{eq:gluon_intermediate}, of which we supply an example in Appendix~\ref{app:gluonportal}, both $f_a$ and $y_\psi$ are bounded. This implies a (somewhat model-dependent) lower bound on $f_{\heta}$ and therefore a lower bound on the lifetime of $\heta$:
\begin{equation}\label{eq:gluonctaumin_bulk}
c\tau \gtrsim 7\,\mathrm{cm}\times \left(\frac{m_a}{500\,\mathrm{GeV}}\right)^{4} \times \left(\frac{5\;\mathrm{GeV}}{m_{\heta}}\right)^{7}.
\end{equation}
Model points that violate this constraint likely have colored fermions with mass $\lesssim 2$ TeV, which is in tension with LHC constraints, as explained in Appendix~\ref{app:gluonportal}. The estimate in \eqref{eq:gluonctaumin_bulk} can be accessed through the function \verb+ctau_min(m)+ in the python package.

\begin{figure}[t]
\includegraphics[width=0.4\textwidth]{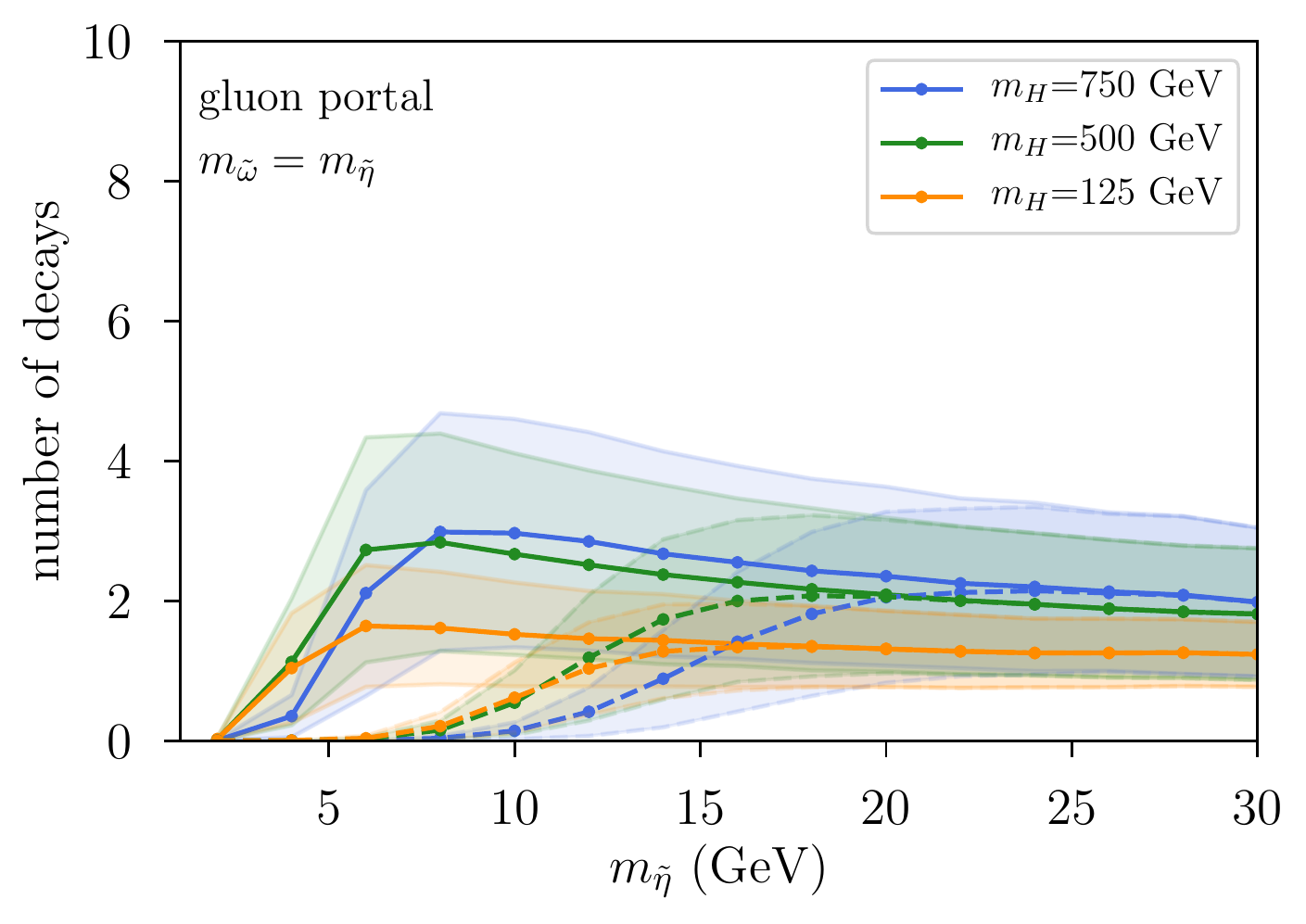}\hfill
\includegraphics[width=0.4\textwidth]{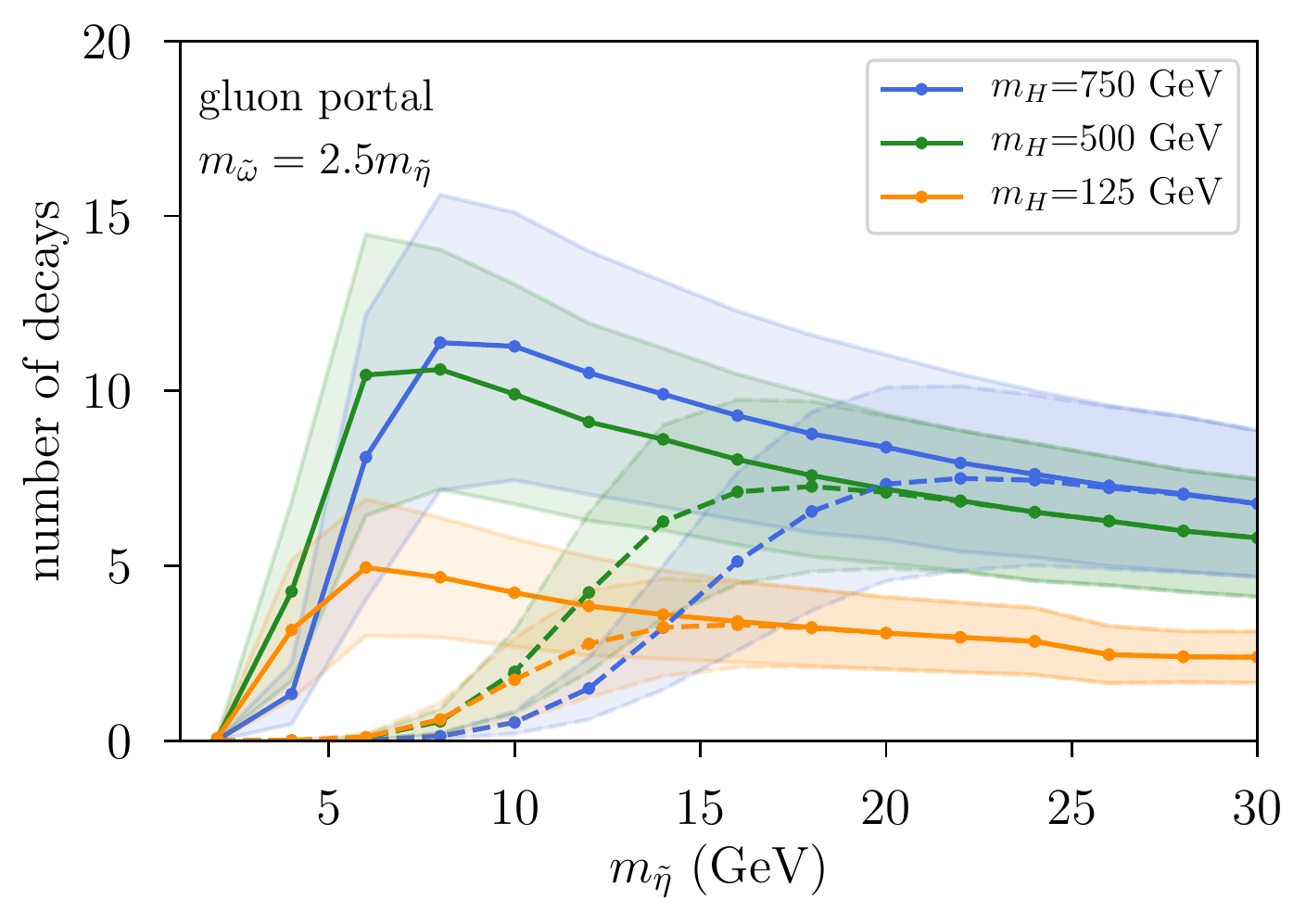}
\caption{Estimated number of $\heta$ mesons with $p_T>5$ GeV and $|\eta|<2.4$ that decay in the tracker, saturating the lower bound on a theoretically well motivated $c\tau$ in \eqref{eq:gluonctaumin_bulk}. Bands represent the $\pm$1 standard deviation of the multiplicity, to indicate the event-by-event variation. The solid (dashed) lines indicate decays satisfying $L_{xy}<$ 1 m ($L_{xy}<$ 1 mm).\label{fig:gluonportal_mult} }
\end{figure}

\begin{figure}[t]
\includegraphics[width=0.475\textwidth]{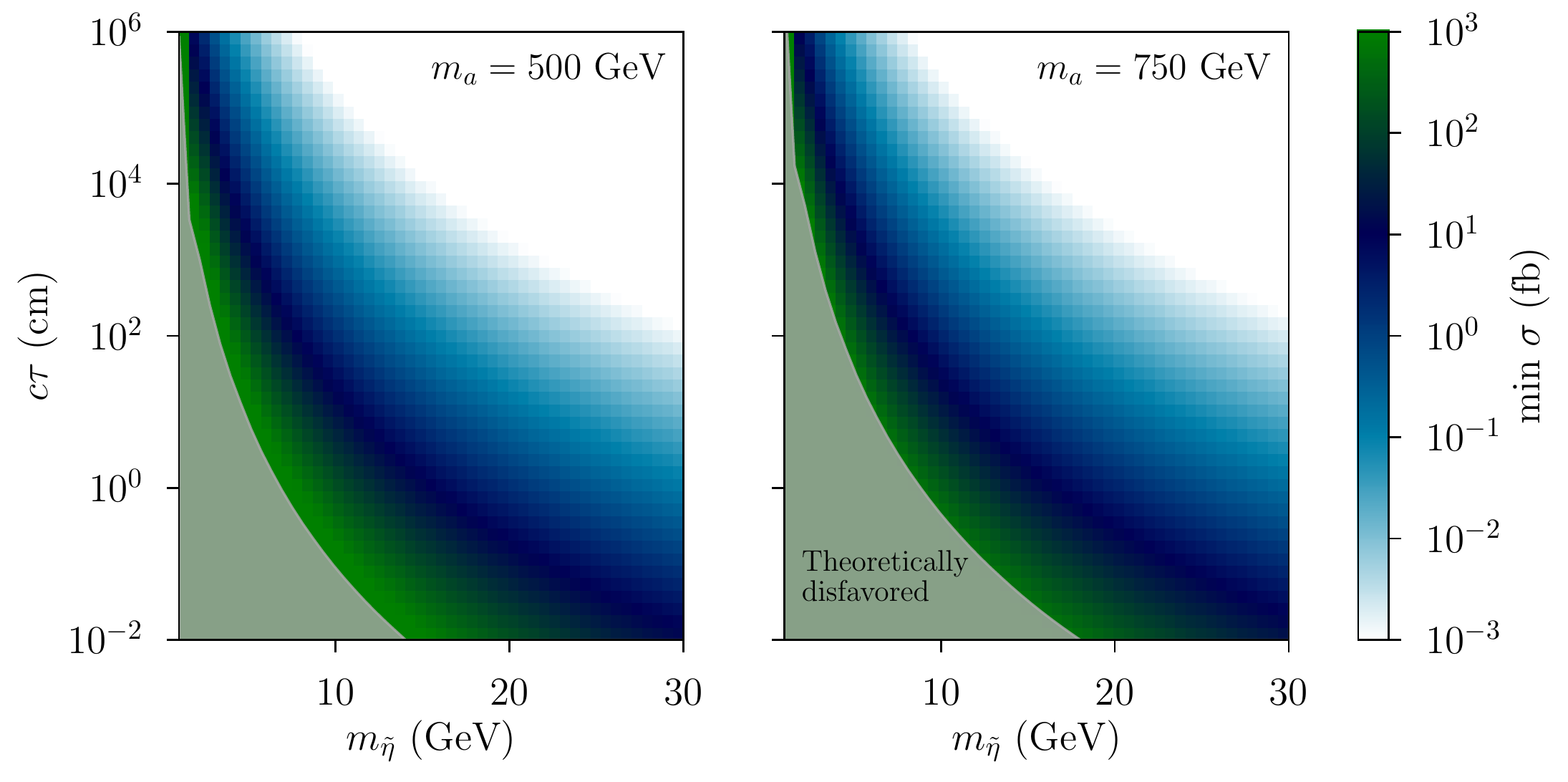}
\caption{Approximate theoretical lower bound on the production cross section as a function of the VDP lifetime in the minimal scenario where the dark shower production and decay both occur through the same gluon portal interaction, as defined in (\ref{eq:gluon_intermediate}). The gray area is theoretically disfavored, as it violates the bound in \eqref{eq:gluonctaumin_bulk}.\label{fig:gluonctauvsxsec}}
\end{figure}

In Fig.~\ref{fig:gluonportal_mult} we plot the corresponding upper bound on the average number of $\heta$ particles with $p_T>5$ GeV and $|\eta|<2.4$ that decay within a given distance from the beamline. We take a benchmark point with $N_c=3$ and $N_\psi=1$, as obtained with PYTHIA 8, and saturate the lower bound in \eqref{eq:gluonctaumin_bulk}. For the 500 GeV and 750 GeV benchmarks we further assume that $a$ is responsible for the production rate as well and thus set $m_a=500$ GeV and $m_a=750$ GeV respectively. For the 125 GeV benchmark, on the other hand, we assume production through an exotic $h\to\psi\bar\psi$ decay, and set $m_a=500$ GeV for the purpose of the lifetime estimates.\footnote{Strictly speaking, we are here neglecting the irreducible production mechanism through $gg\to a\to \psi\bar\psi$. However in most of the parameter space this tends be small compared to the rate that could be contributed by $h\to\psi\bar\psi$ (see Fig.~\ref{fig:gluonctauvsxsec}). } For the solid lines in Fig.~\ref{fig:gluonportal_mult}, we demand that the distance from the decay vertex to the beam line is less than 1 m (\mbox{$L_{xy}<$ 1 m}), which roughly corresponds to decay vertices in the tracker. As in SM QCD, harder dark jets will have a higher particle multiplicity.
 We indeed see that the number of vertices increases with $m_H$ when all other parameters are held fixed. The multiplicity moreover depends strongly on whether the $\homega\to\heta\heta$ decay channel is open; if this channel is closed, the $\homega$ contributes to missing energy. This is meant to illustrate the large uncertainties associated with the dark sector hadronization, as discussed in Sec.~\ref{sec:showerandhadron}. Finally, we see that the multiplicity sharply turns on for \mbox{$m_{\heta}\gtrsim 5$ GeV}, as a consequence of \eqref{eq:gluonctaumin_bulk}: For \mbox{$m_{\heta}\lesssim 5$ GeV} most $\heta$ decay outside the detector volume, and the dark shower topology therefore does not arise.

With the dashed curves we furthermore indicate the multiplicity when insisting on ``prompt'' decays, which we take to mean decays at transverse distances less than 1 mm ($L_{xy}<$ 1 mm) from the beamline. In this case \mbox{$m_{\heta}\gtrsim 15$ GeV} is needed to produce a dark shower topology. This is encouraging news, as the presence of these relatively heavy dark mesons in dark shower events may offer a good handle for
  jet substructure variables and/or machine learning techniques.
In other words, although the gluon
portal model yields a  hadronic signature that appears especially challenging in the prompt regime,
 hidden valley models with perturbative parton showers appear to have difficulties producing dark shower signatures without also introducing a new mass scale substantially distinct from those in the SM.

For the 500 GeV and the 750 GeV benchmarks, the production and decay are assumed to occur through the same portal. This means that the minimum lifetime in \eqref{eq:gluonctaumin_bulk} can be related to the production cross section of the $a$ mediator.  Concretely, the cross section and $\heta$ lifetime scale as $\sigma \sim 1/f_a^2$ and $c\tau \sim f_a^2/y_\psi^2$ respectively, see Appendix~\ref{app:gluonportal}. This means that it is always possible to enhance the lifetime relative to the production cross section by reducing $y_\psi$.  For a given $c\tau$, the cross section is however bounded from below, as the $y_\psi$ coupling in \eqref{eq:gluon_intermediate} cannot be increased arbitrarily. This lower bound is shown in Fig.~\ref{fig:gluonctauvsxsec}.

\subsection{Photon portal\label{sec:photon_alp_portal}}

In analogy to the gluon portal, the lightest (pseudo)scalar meson in the dark sector may also couple to SM photons through 
\begin{equation}\label{eq:photonportal_bulk}
\mathcal{L}\supset\frac{\alpha}{8\pi}\frac{\heta}{f_{\heta}}  F_{\mu\nu}\tilde F^{\mu\nu}\quad\text{and/or}\quad\frac{\alpha}{8\pi}\frac{\heta}{f_{\heta}}\ F_{\mu\nu} F^{\mu\nu},
\end{equation}
again depending on the CP properties of $\heta$. The phenomenology of this scenario is rather intriguing, as events  would essentially feature an (emerging) jet comprised exclusively out of photons \cite{Ellis:2012zp}. This signature so far has not yet received a great deal of attention, though searches involving trackless jets \cite{CMS-PAS-EXO-17-010} ought to have some sensitivity. It deserves a dedicated treatment, the details of which we defer to future work. 

\begin{figure}
\includegraphics[width=0.4\textwidth]{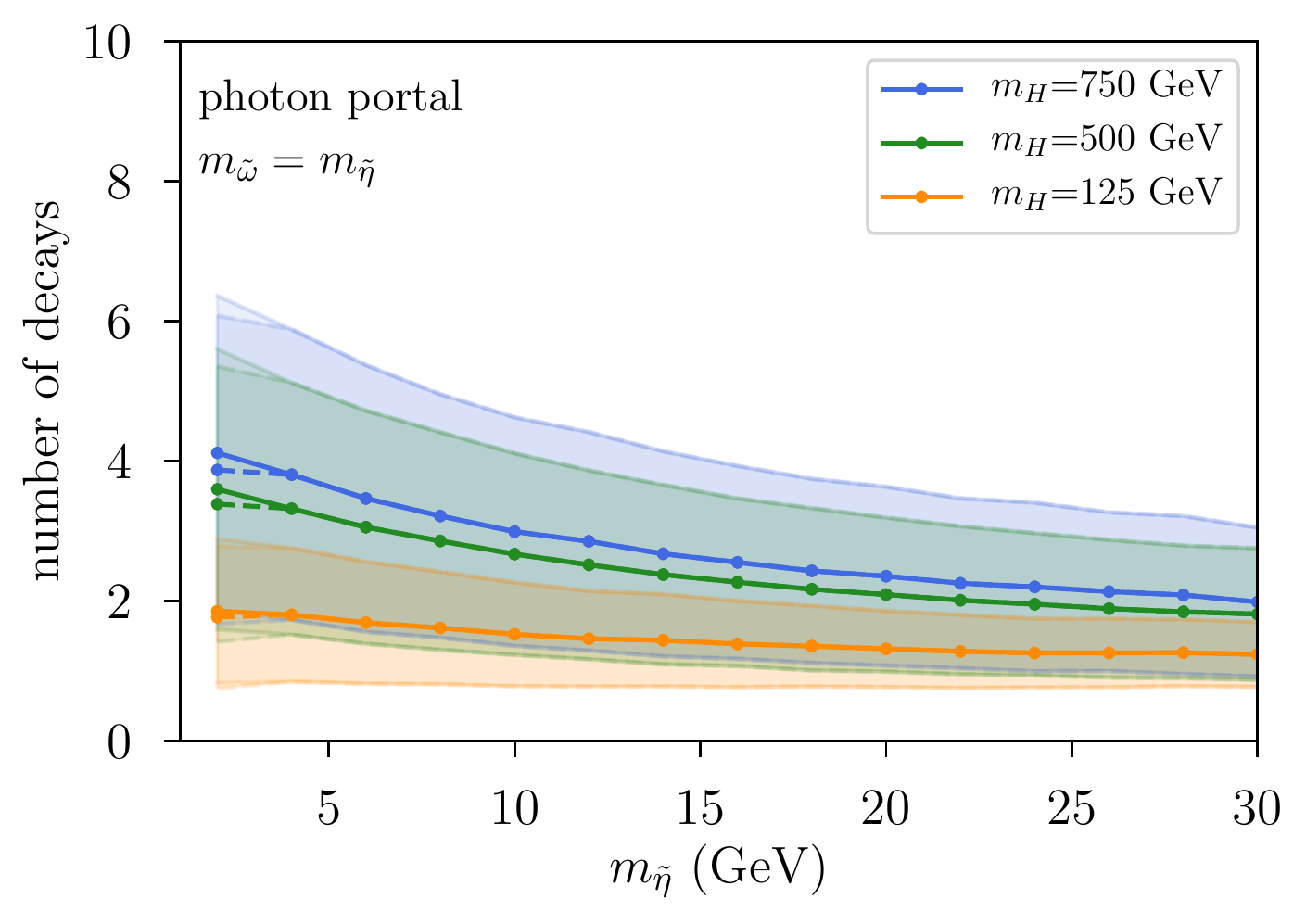}\hfill
\includegraphics[width=0.4\textwidth]{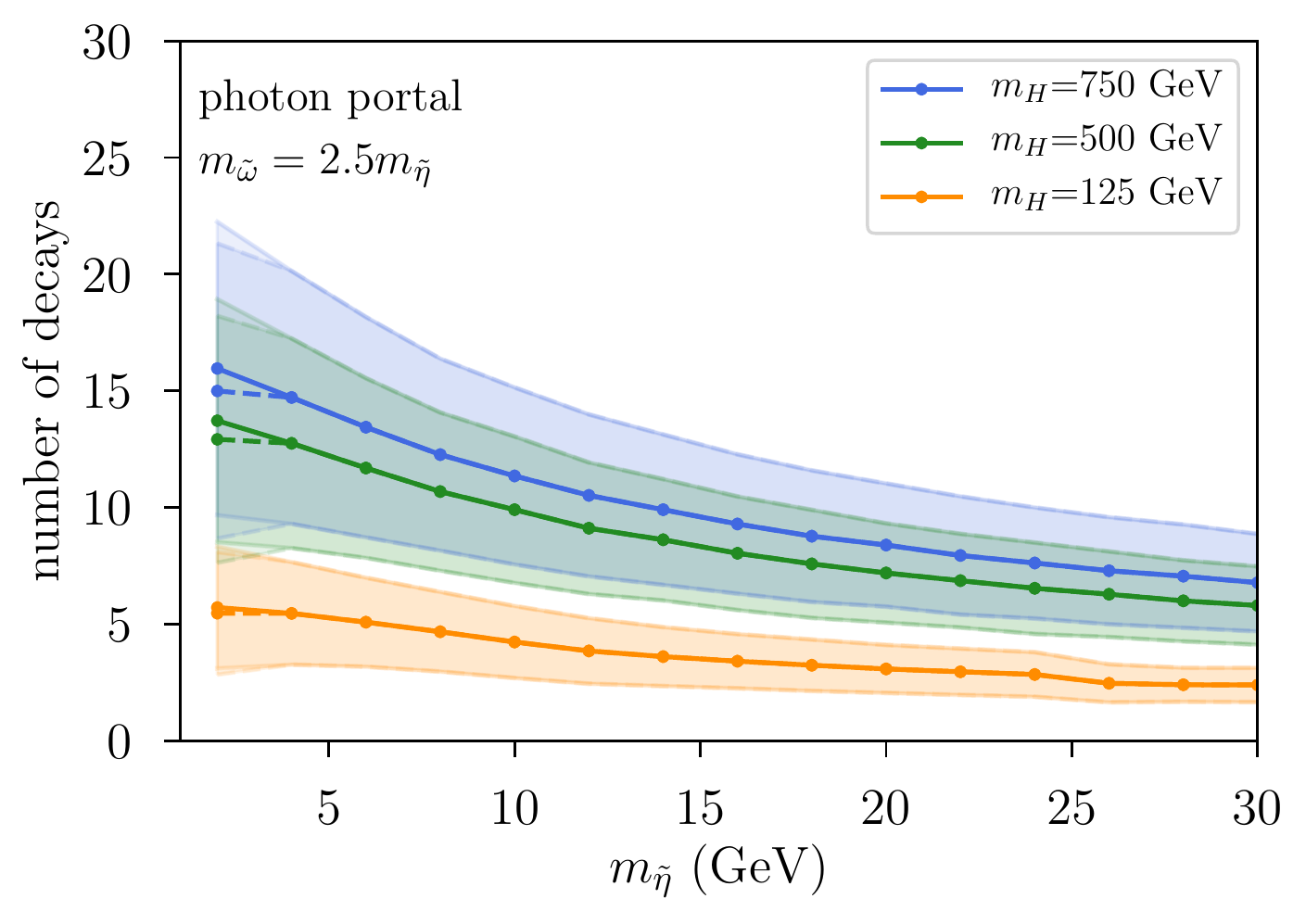}
\caption{Estimated number of $\heta$ mesons with $p_T>5$ GeV and $|\eta|<2.4$ that decay in the tracker, saturating the lower bound on a theoretically well-motivated $c\tau$ in \eqref{eq:photonctauestimate_bulk}. Bands represent the $\pm$1 standard deviation of the multiplicity, to indicate the event-by-event variation. The solid (dashed) lines indicate decays satisfying $L_{xy}<$ 1 m ($L_{xy}<$ 1 mm).\label{fig:photon_portal}}
\end{figure}

We provide an example UV completion of \eqref{eq:photonportal_bulk} analogous to \eqref{eq:gluon_intermediate} in Appendix~\ref{app:photonportal}. In this case, the pseudoscalar $a$ does not have a substantial production cross section, and a separate production portal is needed. The decay width of $\heta$ in terms of $f_{\heta}$ is
\begin{equation}\label{eq:aGGGamma}
\Gamma = \frac{\alpha^2}{256\pi^2}\frac{m_{\heta}^3}{f_{\heta}^2}
\end{equation}  
where $f_{\heta}$ is again a function of the parameters of the UV completion. All UV completions must involve at least one new heavy charged state, which implies an approximate lower bound on $\heta$ lifetime, by accounting for the collider bounds on such charged particles,
\begin{equation}\label{eq:photonctauestimate_bulk}
c\tau \gtrsim 0.1\,\mathrm{cm}\times \left(\frac{1\;\mathrm{GeV}}{m_{\heta}}\right)^{7}.
\end{equation}
As for the gluon portal, the dependence of the bound on $m_{\heta}$ is extremely steep, though in this case  $m_{\heta}$ as low as 1 GeV can be consistent with a dark shower topology, even for prompt decays. This is shown in Fig.~\ref{fig:photon_portal}.

\subsection{Vector portal\label{sec:vector_portal}}

If the hidden sector's hadron spectrum is similar to that of SM QCD, the vector mesons are expected to decay via hidden gauge interactions into lighter hidden sector hadrons. This is particularly the case in $SU(N_c)$  theories with $N_\psi>1$, which, as in QCD, can feature light pseudo-goldstone bosons from chiral symmetry breaking. For $SU(N_c)$ with $N_\psi=1$ (or potentially in more exotic confining theories), however, it is possible for all such channels to be kinematically closed for the lightest vector meson, here modeled with the $\homega$ in the simplified hadron spectrum.  Concretely, in the case of $SU(3)$ with $N_\psi=1$, the lightest hadrons are a pseudoscalar ($\heta$), a scalar ($\hsigma$) and a vector $(\homega)$ meson. The $\heta$ is  not a pseudo-Goldstone boson since the $U(1)$ axial symmetry is anomalous, analogous to the $\eta'$ in the SM, and thus on general grounds one does not expect a large mass hierarchy between $\heta$ and $\homega$ at small $N_c$. 

In this case the lightest vector meson may be the VDP, as it may decay to the SM through a coupling to the SM electromagnetic current \cite{Strassler:2006im}, the weak current \cite{Cheng:2019yai}, or a more general current \cite{Strassler:2006im}. Here we consider the first scenario and show how it can be generated easily by making use of a heavy, kinetically mixed dark photon mediator.  There are also a number of similar models where the dark vector meson instead mixes with a leptophobic $A'$ \cite{Cohen:2015toa,Pierce:2017taw,Cohen:2017pzm,Bernreuther:2019pfb,Bernreuther:2020xus}. The dark sector phenomenology of these models is similar to the simplified model we present here, in the sense that hadronization will produce a mixture of invisible and visible dark mesons, giving rise to a semi-visible jet. The main difference is that with a leptophobic mediator the visible final states will be dominantly hadronic, while with a kinetically mixed dark photon, some $\homega$ will decay to leptons. (The relevant branching ratios are encoded in the python package and can be accessed with the \verb+branching_ratios[channel](m)+ function.) A theory with a leptophobic $A'$ is also only unitary if additional particles are added (see e.g.~\cite{FileviezPerez:2010gw} for an example) or of if the $A'$ couples non-universally to the different SM generations. Given that we use minimality as a theory prior, we therefore opt for a kinetically mixed dark photon mediator, for which these complications are absent. In addition, the phenomenology of the leptophobic $A'$ portal strongly resembles that of the gluon portal and/or Higgs portals with the $\homega\to\heta\heta$ decay closed (Secs.~\ref{sec:gluon_alp_portal} and \ref{sec:Higssmixing}). A kinetically mixed dark photon portal on the other hand produces a distinct signature, in the sense that leptonic and hadronic decays are present with comparable branching ratios. LHCb in particular has good sensitivity for scenarios where light displaced dimuon resonances are realized \cite{Pierce:2017taw}, and a search for this scenario was recently performed \cite{Aaij:2020ikh}.

This scenario is most easily realized by considering an elementary vector boson $A'$ which mixes with the SM hypercharge field strength
\begin{equation}\label{eq:Apdefinition}
\mathcal{L}_{UV}\supset g A'^\mu \bar \psi\gamma_\mu \psi +\frac{\epsilon}{2} B^{\mu\nu}F'_{\mu\nu}
\end{equation}
which yields 
\begin{equation}\label{eq:epseffective}
\mathcal{L}_{IR}\supset \epsilon_{\mathrm{eff}} e J^\mu_{EM} \homega_\mu \quad\mathrm{with}\quad \epsilon_{\mathrm{eff}}\approx g\frac{f_{\homega}m_{\homega}^2}{m_{A'}^2}\epsilon
\end{equation}
in the infrared effective theory, as long as $m_{A'}> m_{\homega}$. 
Here $f_{\homega}$ is the (dimensionless) decay constant of the $\homega$. Lattice studies suggest that $f_{\homega}$ tends to be between 0.1 and 0.3 \cite{DeGrand:2019vbx}; here we take $f_{\homega}=1$ as an estimate which is conservative from the point of view of our lifetime bound.
Eq.~\eqref{eq:epseffective} means that the $\homega$ itself decays like a dark photon, exclusively to charged final SM states.   For $m_{\homega}\gtrsim2 $ GeV, its width is given by 
\begin{equation}
\Gamma_{\homega} = \sum_f \frac{\alpha_{EM}}{3}q_f^2 C_f \epsilon_{\mathrm{eff}}^2 m_{\homega}\left(1+2\frac{m_f^2}{m_{\homega}}\right)\sqrt{1-\frac{4 m_f^2}{m_{\homega}^2}}
\end{equation}
with $\alpha_{EM}$ the electromagnetic fine structure constant, $q_f$ and $m_f$ the electric charge and mass of the SM fermion $f$ and $C_f=1$ ($C_f=3$) for leptons (quarks).  For \mbox{$m_{\homega}\lesssim2$ GeV}, the width is subject to large corrections from mixing with SM hadrons, and it must be extracted from data. In this regime, we use the results obtained in \cite{Buschmann:2015awa} (see also \cite{Ilten:2018crw}).

As for the gluon portal, the $A'$ itself may also serve as the mediator which initiates the dark shower, through the process $q\bar q\to A' \to \psi\bar \psi$. Interesting production rates for this signature are possible because for $g\approx 1$ and $\epsilon\approx 10^{-2}$, the $A'$ can primarily decay to the dark sector, thus alleviating the otherwise stringent constraints on high mass dilepton resonances. If the shower is initiated through an exotic Higgs decay, we can take $m_{A'}$ as light as $\sim 20$ GeV and we find
\begin{equation}\label{eq:vectorportalbound1}
c\tau \gtrsim 7\times10^{-3}\,\text{cm} \times \left(\frac{1 \,\text{GeV}}{m_{\homega}}\right)^5.
\end{equation}
If the shower is instead initiated through the decay of the $A'$ itself, $m_{A'}$ must be much larger and the bound is
\begin{equation}\label{eq:vectorportalbound2}
c\tau \gtrsim 16\,\text{cm} \times \left(\frac{1 \,\text{GeV}}{m_{\homega}}\right)^5\times \left(\frac{m_{A'}}{500\;\mathrm{GeV}}\right)^2.
\end{equation}
where $\epsilon$ was taken to saturate the limits from electroweak precision tests (EWPT). Above the $Z$ pole, these limits scale as $\epsilon \sim 1/m_{A'}$, which explains the perhaps counterintuitive scaling in \eqref{eq:vectorportalbound2}.  We refer to Appendix~\ref{app:vectorportal} for more details. 

\begin{figure}[t]
\includegraphics[width=0.4\textwidth]{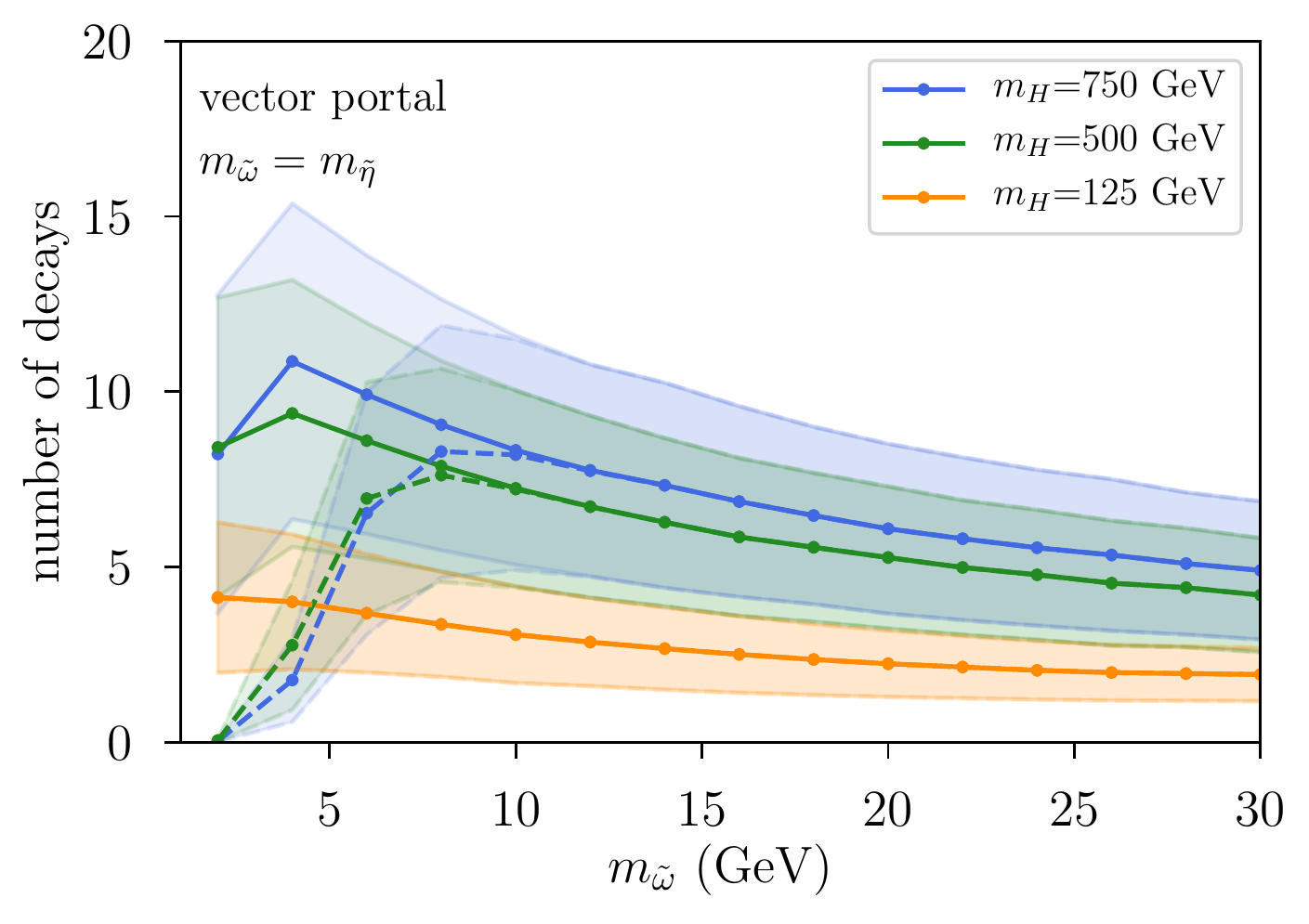}
\caption{Estimated number of $\homega$ mesons with $p_T>5$ GeV and $|\eta|<2.4$ that decay in the tracker, saturating the lower bounds on a theoretically well-motivated $c\tau$ in \eqref{eq:vectorportalbound1} and \eqref{eq:vectorportalbound2}. Bands represent the $\pm$1 standard deviation of the multiplicity, to indicate the event-by-event variation. The solid (dashed) lines indicate decays satisfying $L_{xy}<$ 1 m ($L_{xy}<$ 1 mm).\label{fig:vector_portal}}
\end{figure}

The average vertex multiplicity for the vector portal is shown in Fig.~\ref{fig:vector_portal}, where we identify the $A'$ with the production portal for the 500 GeV and 750 GeV benchmark by setting $m_{A'}=500$ GeV and \mbox{$m_{A'}=750$ GeV} respectively. For the 125 GeV benchmark we assume production through an exotic $h\to \bar\psi\psi$ decay and set $m_{A'}=20$ GeV.\footnote{In this case the $q\bar q \to A' \to \psi\bar\psi$ rate may exceed that of $h\to \bar\psi\psi$, but given the limited phase space of the $A'$ decay, the shower following from direct $A'$ production will have low multiplicity and is likely a very challenging signal.} We only present the case where the $\homega\to\heta\heta$ decay mode is kinematically closed; in the alternative case the $\homega$ is assumed to strongly decay into other dark sector states.  This is a choice made for our simplified benchmarks and not necessarily representative of the specific phenomenology of a particular model, e.g. a realistic $N_\psi=1$ theory, as we discuss in Appendix~\ref{sec:pythia}. The $\heta$ themselves are stable and therefore contribute to missing energy, which means that the vector portal model  generates a semi-visible jet phenomenology \cite{Cohen:2015toa,Cohen:2017pzm} in the regime where the $\homega$ decays promptly.  In the simplified hadronization model we use here, the probability for hadronization to produce an $\heta$ vs.~an $\homega$ is 1:3, as expected from degree-of-freedom counting in the limit where $m_{\heta}\approx m_{\homega}$.   In a more general model this fraction may however be very different, and it is recommended to vary the missing energy fraction when using this model as a benchmark for a semi-visible jet search. This can be accomplished by modifying the \verb+probVector+ flag in PYTHIA 8.

\begin{figure}[t]
\includegraphics[width=0.475\textwidth]{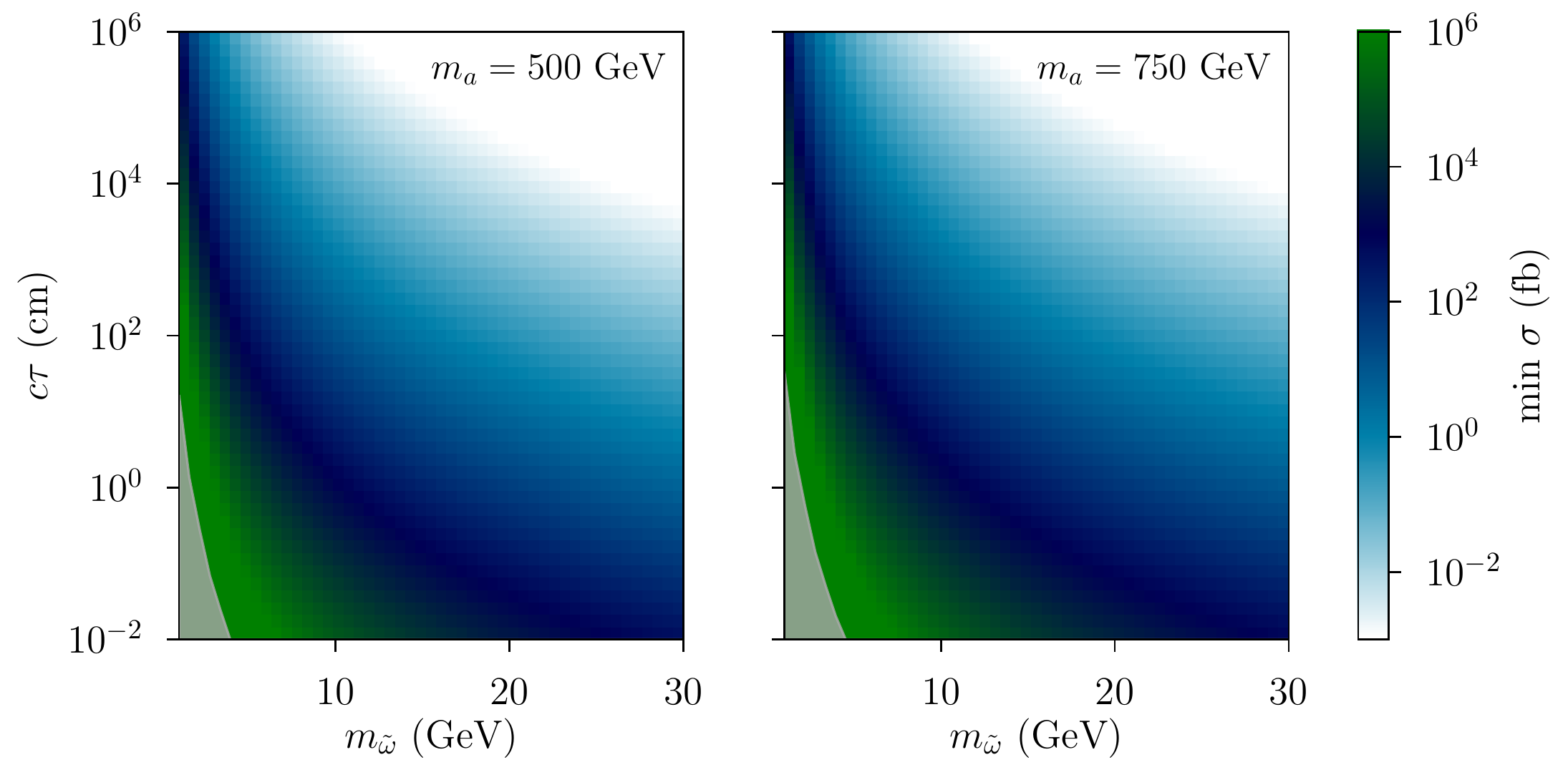}
\caption{Approximate theoretical lower bound on the production cross section as a function of the $\homega$ lifetime, in the minimal scenario where the dark shower production and decay both occur through the same vector portal interaction, (\ref{eq:Apdefinition}). The gray area is theoretically disfavored, as it violates the electroweak precision constraints on $\epsilon$, as discussed Appendix~\ref{app:vectorportal}.\label{fig:vectorctauvsxsec}}
\end{figure}

Finally, if we assume that the $A'$ also serves as the production portal, we can place a lower bound on the cross section as a function of the lifetime of $\homega$, similar to what was done for the gluon portal. Concretely, if we set $m_{\homega}= \tilde \Lambda$ and $g=1$, the mixing parameter $\epsilon$ in \eqref{eq:Apdefinition} controls both the $A'$ cross section and the $\homega$ lifetime. This relation is shown in Fig.~\ref{fig:vectorctauvsxsec}. For a fixed cross section, the lifetime can be longer by taking smaller values of $g$.

\subsection{Higgs portal\label{sec:Higssmixing}}

In the Higgs portal model a light hidden sector meson decays by mixing with the SM Higgs boson. For this to occur there must be a scalar meson in the spectrum which is stable with respect to decays internally to hidden sector. This can occur either because the scalar meson in question is the lightest particle in the dark sector or because no decay channels to lighter dark mesons are kinematically open. The most well-known example of the former is the pure glue hidden valley \cite{Juknevich:2009gg}, which features in some neutral naturalness models \cite{Burdman:2006tz,Craig:2015pha,Craig:2016kue}. In such models the lightest dark sector glueball is a $0^{++}$ state \cite{Morningstar:1999rf} that can mix with the SM Higgs. This effectively occurs through a dimension 6 operator, leading to very long lifetimes \cite{Craig:2015pha}. Other examples are $N_\psi=1$ models at small $N_c$, where the CP-even scalar meson is stable against decays to other hidden sector states \cite{Strassler:2006im},
 theories where the dark sector quarks are scalars rather than fermions, or in dark sectors where CP is violated with an $\mathcal{O}(1)$ amount. 

Here we will work with the same simplified hadronization model as for the other portals, and simply assume that $\heta$ now represents a CP-even scalar rather than a pseudoscalar. Given that the hadronization model was a toy model to begin with, we will not apologize for this somewhat inelegant sleight-of-hand. Concretely, the decay of the $\heta$ is facilitated by the cubic coupling
\begin{equation}\label{eq:etaHH}
\mathcal{L}\supset - \mu \heta H^\dagger H
\end{equation}
that generates $\heta$-Higgs mixing after electroweak symmetry breaking. While \eqref{eq:etaHH} is a relevant coupling, a UV completion is still needed, since $\heta$ is a composite particle. In Appendix~\ref{app:higgsportal} we illustrate how a plausible UV completion will also generate the quartic coupling 
\begin{equation}\label{eq:eta2HH}
\mathcal{L}\supset - \lambda \heta^2 H^\dagger H,
\end{equation}
which generates a contribution to $m_{\heta}$ after electroweak symmetry breaking.  In the absence of tree-level fine tuning, this new contribution cannot exceed $m_{\heta}$, which in turn bounds the size of the cubic interaction in \eqref{eq:etaHH}. This connection is model-dependent however, and in Appendix~\ref{app:higgsportal} we present what we consider the most ``optimal'' scenario, in the sense that it minimizes $c\tau$ without relying on fine-tuning or large anomalous dimensions. (The latter are a priori allowed, but as discussed in Sec.~\ref{sec:showerandhadron}, they are expected to violate important underlying assumptions in the PYTHIA hidden valley module.) The minimum lifetime scales roughly as
\begin{equation}\label{eq:higgs_ctau_bulk}
c\tau \gtrsim 3\,\mathrm{cm}\times \left(\frac{5\;\mathrm{GeV}}{m_{\heta}}\right)^{7}
\end{equation}
though the various kinematical thresholds can provide appreciable corrections to this simple power law depending on the meson mass. These threshholds are responsible for the mild kinks in Fig.~\ref{fig:ctau}. 

\begin{figure}
\includegraphics[width=0.4\textwidth]{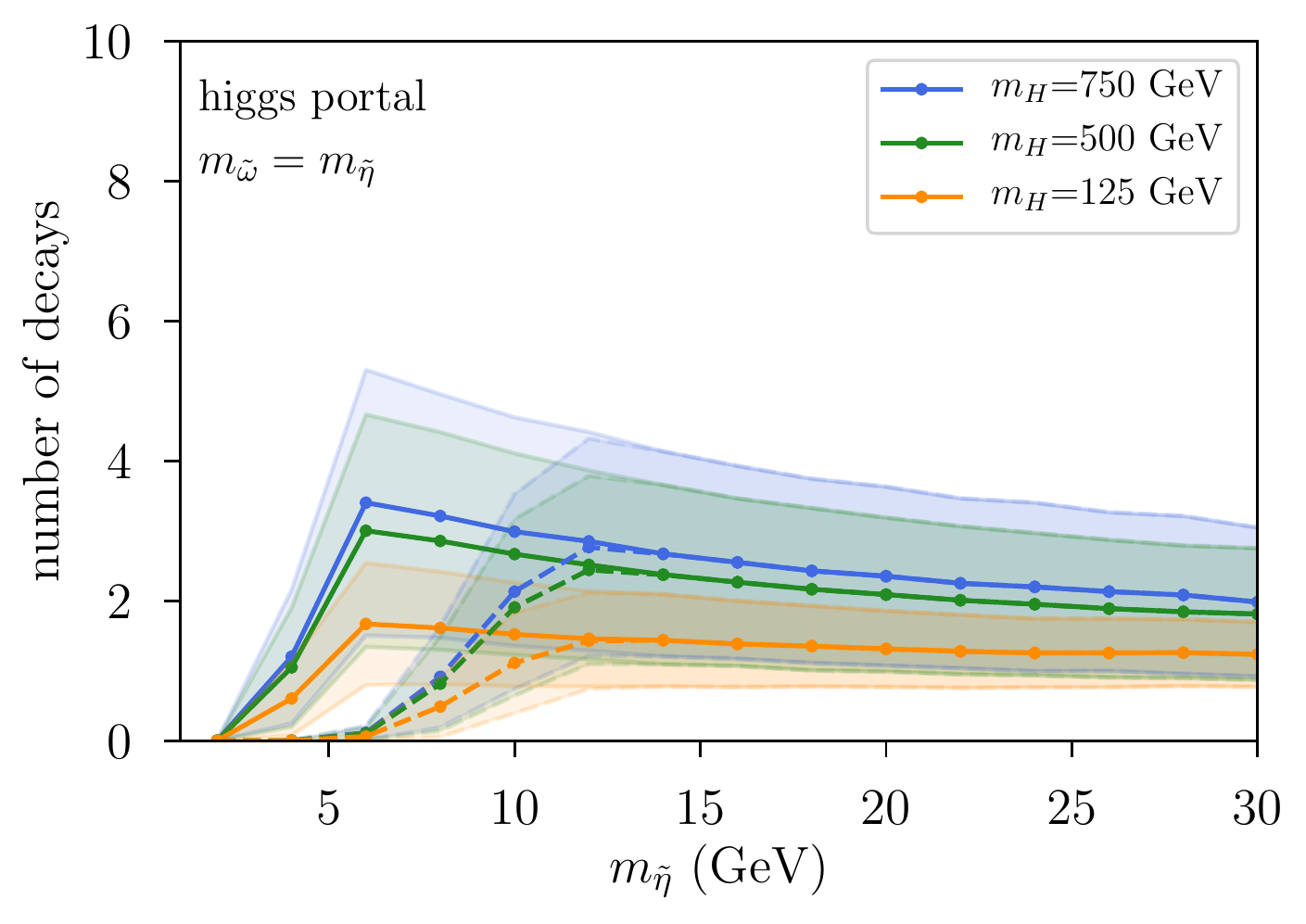}\hfill
\includegraphics[width=0.4\textwidth]{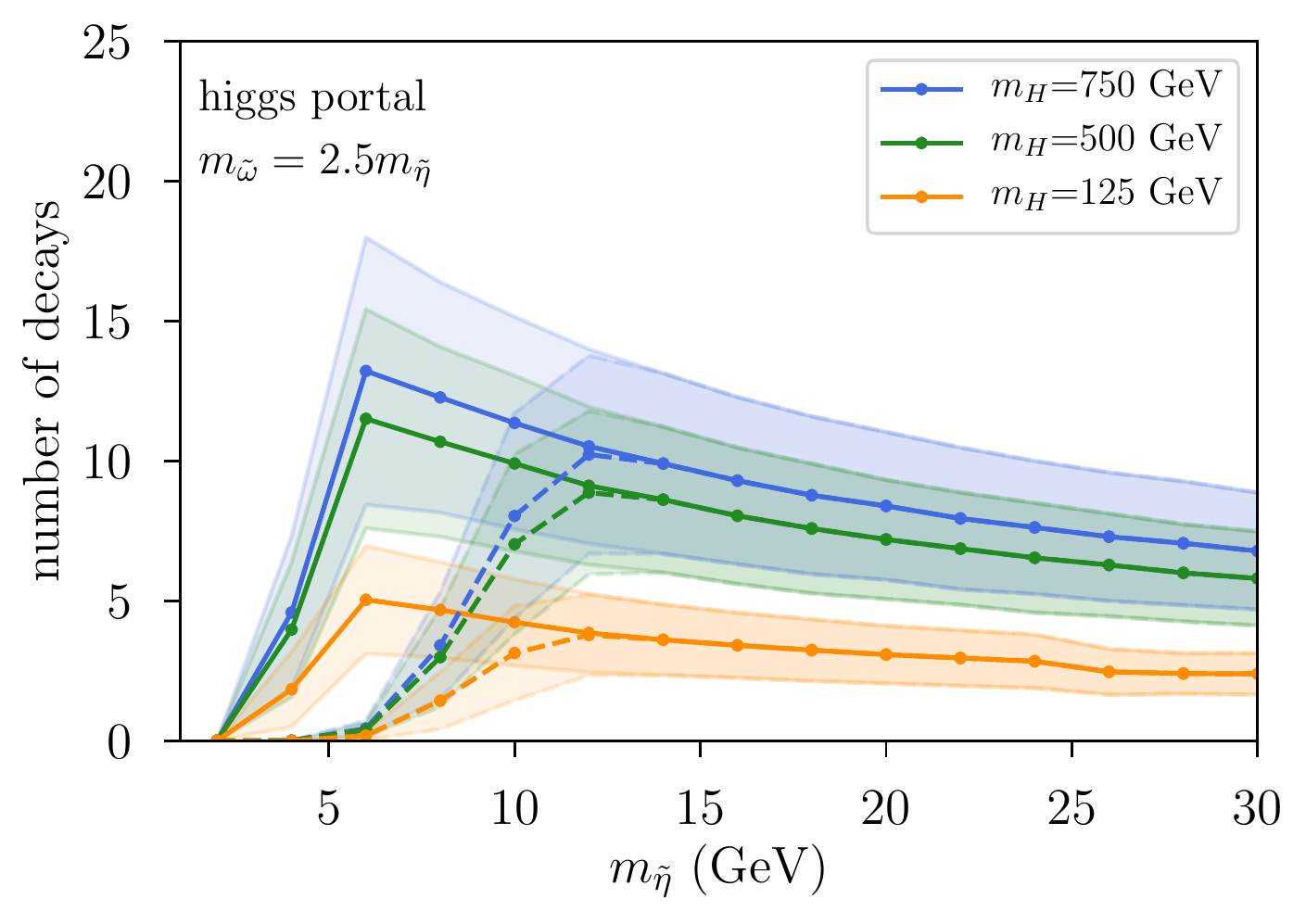}
\caption{Number of $\heta$ mesons with $p_T>5$ GeV and $|\eta|<2.4$ that decay in the tracker, saturating the lower bound on a theoretically well-motivated $c\tau$ in \eqref{eq:higgs_ctau_bulk}.  (See Appendix~\ref{app:higgsportal}). Bands represent the $\pm$1 standard deviation of the multiplicity, to indicate the event-by-event variation. The solid (dashed) lines indicate decays satisfying $L_{xy}<$ 1 m ($L_{xy}<$ 1 mm).\label{fig:higgs_portal}}
\end{figure}

The corresponding vertex multiplicity is shown in Fig.~\ref{fig:higgs_portal}, where for the 500 GeV and 750 GeV benchmarks we must assume an additional mediator to initiate the shower. The $a$ or $A'$ particles in Sec.~\ref{sec:gluon_alp_portal} and Sec.~\ref{sec:vector_portal} are possible candidates for this. As was the case for the gluon portal, we find that a dark shower topology is less plausible for $m_{\heta}\lesssim 5$ GeV, though in the Higgs portal model the bound is motivated by (tree-level) fine-tuning rather than by experimental constraints on new colored particles. This implies that the $\heta\to \tau\tau$ and $\heta\to c\bar c$ channels are always open, while $\heta\to b\bar b$ is open in most of the relevant parameter space. In this sense, the Higgs portal benchmark will always produce a dark shower that is enriched in heavy flavors. The importance of heavy flavor final states was emphasized early on in the development of hidden valley models, along with several analysis strategies \cite{Strassler:2008fv}.

\subsection{Dark photon portal\label{sec:lightDP}}

\begin{figure}
\includegraphics[width=0.4\textwidth]{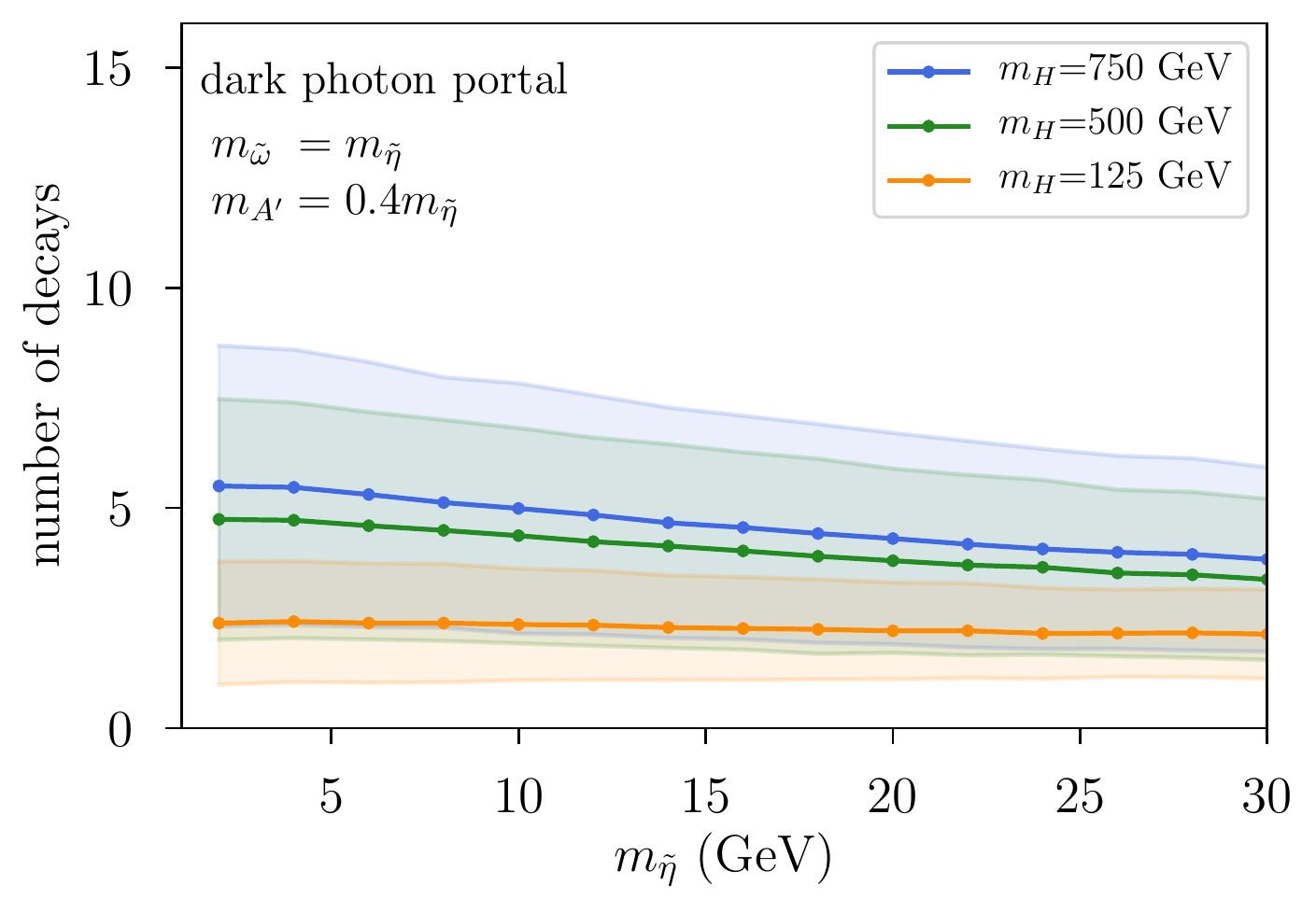}\hfill
\includegraphics[width=0.4\textwidth]{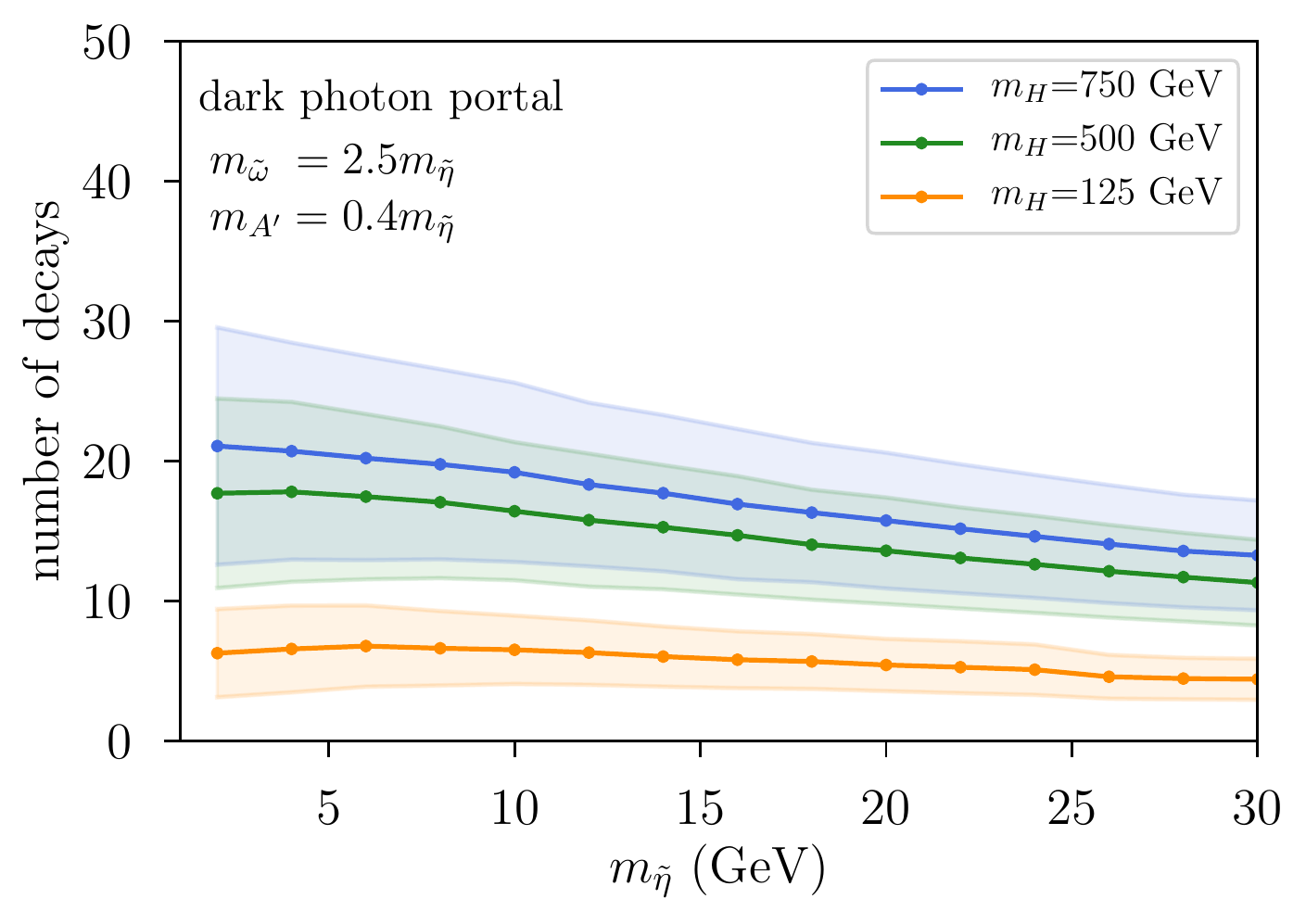}
\caption{Number of dark photons  $A'$ with $p_T>5$ GeV and $|\eta|<2.4$. Bands represent the $\pm$1 standard deviation of the multiplicity, to indicate the event-by-event variation. \label{fig:darkphoton_portal}}
\end{figure}

The dark photon portal is similar to the vector portal in terms of the matter content of the theory; however we are now interested in the case where $m_{A'}<2 m_{\heta}$. We further assume that the $\heta$ couples to the dark photon through an anomaly, similar to the $\pi^0\gamma\gamma$ and $\eta \gamma\gamma$ couplings in the SM. Specifically, we write
\begin{equation}
\mathcal{L}\supset - \frac{1}{f_{\heta}} \heta F'^{\mu\nu} \tilde F'_{\mu\nu} - \frac{\epsilon}{2} F'^{\mu\nu} F_{\mu\nu}.
\end{equation}
The decay of the $\heta$ therefore occurs in two steps: the \mbox{$\heta\to A'A'$} decay can always occur promptly, while the \mbox{$A'\to \text{SM}$} decay is only constrained by the experimental upper bound on the mixing parameter $\epsilon$. For \mbox{$m_{A'}\gtrsim 10$ MeV}, prompt decays of the $A'$ are currently allowed, see e.g.~\cite{Ilten:2018crw}. Up to phase space effects, the branching ratios of the $A'$ are roughly proportional to the electric charge of the final states squared, which means that this type of dark shower is relatively rich in leptons.  Because of the additional step in the decay chain, the typical vertex multiplicity is also roughly a factor of two higher than for the other portals. We assume that the $\homega$ promptly decays to a pair of $\heta$ mesons when possible, and is detector-stable otherwise. The latter case is strictly speaking not fully self-consistent, as the $\homega$ would also decay to the visible sector through the (off-shell) $A'$, which was the subject of Sec.~\ref{sec:vector_portal}. On the other hand, the PYTHIA hadronization model neglects a number of hidden sector mesons that might very well be invisible if they were present. Here we view the $\homega$ as loosely representing these missing states in the hadronization model, to avoid the erroneous impression the dark photon portal would always imply the absence of invisible states. The features discussed above are illustrated in the multiplicity plots in Fig.~\ref{fig:darkphoton_portal}.

The implementation of the dark photon portal decay modes in the python package is the same as that for the vector portal, as described in Sec.~\ref{sec:vector_portal} and Appendix~\ref{app:vectorportal}. The minimum experimentally allowed lifetime of $\heta$ is specified in the \verb+ctau_min(m)+ function, with \verb+m+ representing $m_{\heta}$. Here we fixed $\epsilon=3\times 10^{-4}$, roughly saturating the existing BaBar \cite{Lees:2014xha} and LHCb \cite{Aaij:2019bvg} limits for $m_{A'}\lesssim 10$ GeV. The dark photon mass is specified by the parameter $\xi_{A'}\equiv m_{A'}/m_{\heta}$, represented by the optional \verb+xi_Ap+ flag in \verb+ctau_min+, for which the default is set to 0.4. This parameter must be set to a value smaller than 0.5 for the $\heta\to A'A'$ mode to be kinematically allowed.

\section{Discussion\label{sec:discussion}}

We present a set of five benchmark hidden valley models featuring a perturbative parton shower, which we believe constitute a good basis for a systematic experimental exploration of this space of models. Our motivation is phenomenological, in the sense that we hope to span a broad space of possible dark shower signatures, though given the vast landscape of possibilities, we must necessarily inject some theory priors in order to do so economically.  To this end we choose \emph{minimality} as a guiding principle, restricting ourselves to models with a single visibly-decaying particle (VDP) that can decay through one of a few ``minimal'' decay portals.  These well-motivated portal operators give rise to a diverse suite of experimental signatures, summarized in  Tab.~\ref{tab:portals}, and ensure that the branching ratios of the VDP are fully predicted as a function of its mass.  Other, less minimal decay portals are  possible beyond the scenarios treated here, e.g.~leptophobic vector portals \cite{Cohen:2015toa,Pierce:2017taw,Cohen:2017pzm,Bernreuther:2019pfb,Bernreuther:2020xus}.  In many cases such models will produce similar phenomenological signatures to the benchmarks considered here; we leave a detailed assessment of the coverage of the dark shower signature space to future work.
For the production portal we use $s$-channel rather than $t$-channel mechanisms, in order to minimize signal dependence on the (model-specific) accompanying hard SM objects inevitable in $t$-channel production.   Our aim here is to provide benchmarks that help facilitate the development of searches sensitive to the dark shower signature itself, as such searches are necessary for an inclusive and comprehensive discovery program.  For showering and hadronization within the dark sector, these benchmarks rely on PYTHIA's Hidden Valley module.  

 Given these minimal theoretical priors, we can make the following observation:  there are loose lower bounds on the lifetime of the VDP as a function of its mass, depending in detail on the choice of decay portal.  These lower bounds are not rigorous exclusions, but rather statements about the kinds of lifetimes that are readily  achievable in models that generate the portal operator allowing the VDP to decay.  For Higgs and gluon decay portals in particular, dark shower topologies are only feasible for VDPs with mass larger than roughly 5 GeV.   We emphasize that the lifetime bounds that we find here assume small anomalous dimensions for the operators describing dark hadrons, and are thus only applicable for the perturbative shower models we consider.  One way to relax these lifetime bounds is to consider hidden sectors with large 't Hooft couplings, where hadrons may in some cases be described by operators with large anomalous dimensions; however, in those regimes, we expect the energy flow within the dark sector to be substantially different from the perturbative parton shower modeled here.  
 
  All models admit production through exotic Higgs decays.  This production mode is model-independent, in the sense that the properties of the Higgs mediator are now well-understood, and the overall center-of-mass energy is fixed.  We also consider a more general case, with a new, heavier $s$-channel mediator for each of the portals. For gluon and vector portal models, the same mediator can also be responsible for the VDP's decay in a relatively minimal fashion.  When this additional assumption is made, we derive an approximate lower bound on the production cross section as a function of the lifetime of the dark meson.

\section*{Acknowledgments}
We thank Matthew Citron, Nathaniel Craig, Caterina Doglioni,  Marat Freytsis, Dorota Grabowska, Simone Pagan Griso, Myeonghun  Park, Diego Redigolo, Ennio Salvioni, Yotam Soreq, Matthew Strassler, Yuhsin Tsai and Mengchao  Zhang for useful discussions. We are particularly grateful to Nathaniel Craig for reminding us of the importance of lifetime considerations for the Higgs portal model and to Matthew Citron, Caterina Doglioni, Marat Freytsis, Ennio Salvioni and Matthew Strassler for valuable comments on the manuscript. We additionally thank Matthew Strassler for pointing out an error in version one of this work, related to our interpretation of the pythia hidden valley hadronization module with respect to its treatment of pseudo-goldstone bosons.
  This research was supported by a grant from the United States - Israel Binational Science Foundation (BSF), Jerusalem, Israel and DOE CAREER grant DE-SC0017840. The work of SK was in part supported by DOE grant DE-SC0009988. This research used resources of the National Energy Research Scientific Computing Center (NERSC), a U.S. Department of Energy Office of Science User Facility operated under Contract No. DE-AC02-05CH11231. This work was performed in part at the Aspen Center for Physics, which is supported by National Science Foundation grant PHY-1607611.

\appendix

\section{Hadronization in PYTHIA's Hidden Valley module}
\label{sec:pythia}

For Monte Carlo generation, we rely on the PYTHIA 8 hidden valley module \cite{Carloni:2010tw,Carloni:2011kk}, which restricts the hadronization model we can consider. 
Concretely, we consider a confining $SU(N_c)$  gauge theory with $N_\psi$ vectorlike flavors, which are all degenerate in mass. 

In a realistic $N_\psi=1$ theory, the lightest hadrons are a pseudoscalar ($\heta$), a scalar ($\hsigma$) and a vector $(\homega)$ meson. The mass ratio between the $\heta$ and the $\homega$ famously scales as $m_{\heta}/m_{\homega}\sim 1/\sqrt{N_c}$ \cite{Witten:1979vv}. The stripped-down hadronization model used in the PYTHIA 8 hidden valley module with $N_\psi =1$ does not include the $\hsigma$ resonance, which is expected to be narrow for $N_\psi=1$ and small $N_c$:  Lattice calculations have shown that $m_{\hsigma}/m_{\heta}\approx1.5$  for $N_c=3$ \cite{Farchioni:2007dw}, which means that the $\tilde \sigma$ is stable in this theory. By virtue of the $\hsigma$'s higher mass, it is expected to be produced at somewhat lower rates than the $\heta$; however, in neglecting the $\hsigma$, PYTHIA 8 is likely overpredicting the number of $\heta$ and $\homega$ mesons compared to predictions from a full realization of one-flavor QCD.  
Finally, dark baryons are also neglected in the dark sector hadronization.

In the $N_\psi=2$ case, 
the spectrum now additionally features a triplet of pseudo-goldstone bosons ($\hpi^{0},\hpi^{\pm}$) that carry the dark sector's equivalent of isospin. In such theories, the $\hpi$ masses are controlled by the dark quark masses and are therefore can be parametrically lighter than those of the other mesons. All dark mesons are moreover electrically neutral, and the $0,\pm$ superscripts refer to the charges of the dark pions under the dark sector analogue of the electromagnetic $U(1)$ symmetry. This $U(1)$ may be a global or a gauge symmetry, and in the former case it may be unbroken. 
In the $N_\psi>1$ scenario, the PYTHIA 8 hidden valley fragmentation model neglects the hidden sector analogues of the Standard Model $\eta$ and $\sigma$ mesons, which is reasonable as long as dark pions are parametrically lighter. For $N_\psi>1$, PYTHIA's spin-one mesons should similarly be understood to be dark isospin adjoints, analogous to the SM $\rho$, rather than dark isospin singlets.  


Our benchmark hadronization models consider the production of dark flavor singlet meson states, $\heta$ and $\homega$, and assume that the decay $\homega\to \heta\heta$ occurs promptly when it is kinematically accessible.   This is a toy model, but nonetheless we expect this particular simple toy model to be useful.
In a realistic $N_\psi=1$ model, the $\heta$ and $\homega$ are indeed flavor singlets, but the $\omega\to\heta\heta$ decay channel is forbidden by Bose symmetry and angular momentum conservation.  The $\hsigma$, not included in the PYTHIA module, does decay to a pair of $\heta$ when kinematically allowed, and will then contribute to the final $\heta$ multiplicity.  This occurs in particular in models with $N_c\gg1$. However our primary motivation for allowing the $\omega\to\heta\heta$ decay  to proceed in the toy model is to provide a way to vary the $\heta$ multiplicity by dialing the mass ratio between $m_{\heta}/m_{\homega}$.    
Meanwhile in $N_\psi =2$ models where all flavor symmetries are explicitly broken (e.g.~by a combination of quark mass matrices and portal couplings), vector mesons will decay to pairs of VDP dark pions when kinematically allowed.  In this regime the toy model with its single species of VDP realizes a similar phenomenology while avoiding detailed hard-coding of model-dependent meson decays.

\begin{figure}[t]
\includegraphics[width=0.4\textwidth]{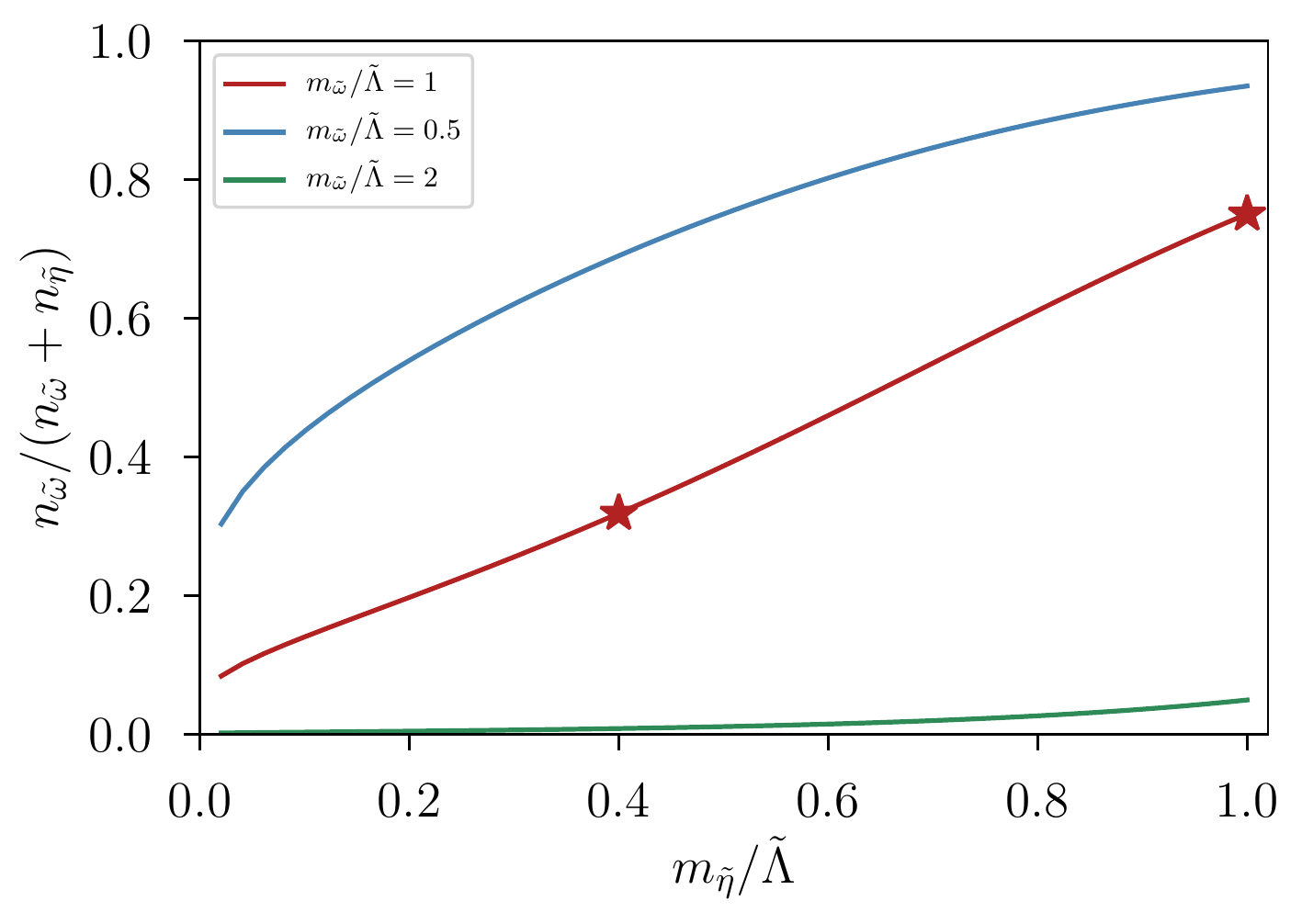}
\caption{Fraction of dark mesons with spin-one in the simplified hidden valley hadronization model used here. The stars indicate the benchmark points selected in this paper. The gray shaded area indicates that this toy fragmentation model is a not plausible description in the $m_{\heta}/\hLambda\to 0$ limit. \label{fig:hadronizationplot}}
\end{figure}

Now we turn to the toy fragmentation scheme to set the fraction of dark vector mesons produced in the dark parton shower. For degenerate spin-one $\homega$ and spin-zero $\heta$, the ratio of their average multiplicities should follow simply from counting spin states, giving $3:1$. However, the difference in the masses of $\homega$ and $\heta$ should be reflected in the fragmentation function $f(z)$. Loosely inspired by the Lund fragmentation model \cite{Andersson:1983jt,Andersson:1983ia}, we assume a fragmentation function of the form
\begin{equation}
\label{e.fragmentation_reduced}
f(z,m) = N\frac{1}{z}\mathrm{exp}^{-\frac{m^{2}}{z\hLambda^2}},
\end{equation}
with $N$ a normalization constant, $m$ the mass of the meson in question and $z$ the momentum fraction of the meson relative to the momentum of the parton. This toy model is meant as a simplified ansatz for the regime where the meson masses are not parametrically below the confinement scale. The average multiplicities are then
\begin{align}
n_{\homega} &=  3\int^{1}_{0} f(z,m_{\homega}) \mathrm{d}z=3NE_1\left(\frac{m_{\homega}^2}{\hLambda^2}\right)\\
n_{\heta} &=  \int^{1}_{0} f(z,m_{\heta}) \mathrm{d}z=N E_1\left(\frac{m_{\heta}^2}{\hLambda^2}\right),
\end{align}
with $E_1(x)$ the exponential integral function. The factor of 3 accounts for the degrees of freedom in the spin-one vector meson. The fraction of spin-one mesons that are produced is then
\begin{equation}
\frac{n_{\homega}}{n_{\homega}+n_{\heta}}=\frac{3E_1\left(\frac{m_{\homega}^2}{\hLambda^2}\right)}{3E_1\left(\frac{m_{\homega}^2}{\hLambda^2}\right)+E_1\left(\frac{m_{\heta}^2}{\hLambda^2}\right)}.
\end{equation}
This ratio is shown in Fig.~\ref{fig:hadronizationplot} and is set with the \verb+probVector+ flag in the PYTHIA 8 hidden valley module. In this paper we use the benchmark points $\hLambda=m_{\homega}=m_{\heta}$ and $\hLambda=m_{\homega}=2.5 m_{\heta}$, which  correspond respectively to an $\homega$ hadronization fraction of  0.75 and 0.32. 

\section{Details on UV completions and lifetime estimates}
\label{sec:lifetime}

In the discussion below, we estimate the minimum feasible values of $c\tau$ for hidden sector hadrons decaying through various portals into the SM.  

\subsection{Gluon portal\label{app:gluonportal}} 
As an example UV completion of the gluon portal, we consider an elementary pseudo-scalar particle ($a$) and a set of $N_Q$ vectorlike color triplet fermions ($Q$) whose interactions are given by
\begin{align}\label{eq:gluon_UVcompletion}
\mathcal{L}_{UV}\supset& - M_Q \sum_{i=1}^{N_Q}\bar Q_i Q_i - m_\psi \bar \psi \psi -\frac{1}{2}m_a^2a^2\nonumber\\
&-i y_Q a \bar Q \gamma_5 Q - i y_\psi a \bar \psi \gamma_5 \psi +\text{h.c.},
\end{align}
with  $\bar\psi$, $\psi$ the constituent ``quarks'' in the hidden sector. Upon integrating out the $Q_i$ we obtain the IR effective action,
\begin{equation}\label{eq:gluon_portal_intermediate}
\mathcal{L}_{IR}\supset-\frac{1}{2}m_a^2 a^2 -\frac{ \alpha_s}{8\pi}\frac{y_Q N_Q}{M_Q} a G\tilde G  - i y_\psi a \bar \psi \gamma_5 \psi +\text{h.c.},
\end{equation}
with $\tilde G^{\mu\nu}\equiv \frac{1}{2}\epsilon^{\mu\nu\rho\sigma}G_{\rho\sigma}$.  We can further perform the mapping $\bar\psi \gamma^5\psi\to \hLambda F_{\heta} \heta$ with $F_{\heta}$ and $\hLambda$ respectively $\heta$'s strong decay constant and the dark sector confinement scale. Further integrating out $a$ as well, we finally obtain the effective interaction
\begin{align}
\mathcal{L}_{IR}&\supset \frac{ \alpha_s}{8\pi}\frac{1}{f_{\heta}}\heta G\tilde G,
\end{align}
with $f_{\heta}$ the effective decay constant of $\heta$, given by
\begin{align}\label{eq:fheta_indent}
\frac{1}{f_{\heta}}= \frac{y_Qy_\psi N_Q \hLambda F_{\heta}}{m_a^2 M_Q}.
\end{align}

To estimate the minimum realistic value of $f_{\heta}$, it suffices to insert numbers for the various parameters in \eqref{eq:fheta_indent}. This necessarily involves some degree of taste, in particular how large one is willing to take $y_Q$, $y_\psi$, and $N_Q$. Here we settle on $y_Q=y_\psi=3$ and $N_Q=1$. We further take $F_{\heta}\approx\hLambda\approx m_{\heta}$ and $M_Q\gtrsim 2$ TeV, where the latter is (conservatively) motivated by collider constraints on heavy colored fermions (see e.g.~\cite{Aaboud:2018ifs}). Different choices are of course possible, and the reader can rescale our results to meet their own tastes and theory priors. With our choices, we find
\begin{equation}\label{eq:effectivedecay}
f_{\heta}\gtrsim 2\times 10^6\, \mathrm{GeV}\times \left(\frac{m_a}{500\,\mathrm{GeV}}\right)^2 \times \left(\frac{5\;\mathrm{GeV}}{m_{\heta}}\right)^2
\end{equation}
which corresponds to 
\begin{equation}\label{eq:gluonctauestimate}
c\tau \gtrsim 7\,\mathrm{cm}\times \left(\frac{m_a}{500\,\mathrm{GeV}}\right)^{4} \times \left(\frac{5\;\mathrm{GeV}}{m_{\heta}}\right)^{7}.
\end{equation}
This number is to be understood as an order-of-magnitude estimate only, given the many assumptions that were needed to arrive at \eqref{eq:gluonctauestimate}. On the other hand, given the extremely strong dependence of $c\tau$ on $m_{\heta}$, a shift of even one or two orders of magnitude in the lower bound on $c\tau$ would not qualitatively change the minimum value of $m_{\heta}$ at which a dark shower topology is possible.

The estimated bound in \eqref{eq:gluonctauestimate} can be accessed in the python package through the function \verb+ctau_min(m)+, where $m_a$ can be specified with the optional parameter $\verb+m_a+$, for which the default value is 400 GeV. Similarly, the ratio $\xi_\Lambda=\hLambda/m_{\heta}$ can be specified using the optional parameter $\verb+xi_Lambda+$, for which the default value is 1. The parameters $M_Q$, $N_Q$, $y_Q$ and $y_\psi$ as well as the relation $F_{\heta}=\hLambda$ must be modified directly in the source code. Alternatively, bounds for alternate choices can be obtained by a straightforward rescaling, using \eqref{eq:fheta_indent}.

 Finally, we briefly comment on the implementation of the exclusive decay modes for $m_{\heta}<1.84$ GeV in the python package, for which we largely follow Aloni et al.~\cite{Aloni:2018vki}. Concretely, Tab.~\ref{tab:alpbranchingratios} shows the various channels that are included in our python code, with their breakdown in terms of charged vs.~neutral final states. The estimates for the latter are taken from \cite{Aloni:2018vki} when possible and otherwise obtained with chiral perturbation theory, where we neglected any dependence on $m_{\heta}$. 
The charge to neutral ratio of the various final states is a priori important for searches relying on displaced tracks and/or displaced vertices, though as we have seen in \eqref{eq:gluonctauestimate}, for dark shower topologies the inclusive decay in \eqref{eq:aglueglue} is always appropriate. The decay length is implemented in the python package as the function \verb+ctau(m,f)+ where \verb+m+ and \verb+f+ are the mass of the VDP $(m_{\heta})$ and its decay constant  $(f_{\heta})$. The branching ratios can be accessed with the \verb+branching_ratios[channel](m)+ function.

\begin{table}[t]
\begin{tabular}{l|l}
Channels & neutral/charge ratios\\\hline
$\heta\to \gamma\gamma$ &/\\
$\heta\to 3\pi$ &$\Gamma_{\heta\to 3\pi^0}=\frac{3}{2}\Gamma_{\heta\to \pi^+\pi^-\pi^0}$\\
$\heta\to \pi^+\pi^-\gamma$ &/\\
$\heta\to  \pi\pi\eta$ &$\Gamma_{\heta\to 2\pi^0\eta}=\frac{1}{2}\Gamma_{\heta\to \pi^+\pi^-\eta}$\\
$\heta\to  \pi\pi\eta'$ &$\Gamma_{\heta\to 2\pi^0\eta'}=\frac{1}{2}\Gamma_{\heta\to \pi^+\pi^-\eta'}$\\
$\heta\to KK\pi$ &$\Gamma_{\heta\to K^+K^-\pi^0}=\Gamma_{\heta\to K^\pm K_{S,L}\pi^\mp}=\Gamma_{\heta\to K_SK_L\pi^0}$\\
$\heta\to \rho\rho$ &$\Gamma_{\heta\to \rho^0\rho^0}=\frac{1}{2}\Gamma_{\heta\to \rho^+\rho^-}$\\
$\heta\to K^\ast\bar K^\ast$ &$\Gamma_{\heta\to K^{\ast0}\bar K^{\ast0}}=\Gamma_{\heta\to K^{\ast+}K^{\ast-}}$\\
$\heta\to \omega\omega$ &/\\
\end{tabular}
\caption{Implemented decay channels and their breakdown in terms of charged vs neutral final states. \label{tab:alpbranchingratios}}
\end{table}

\subsection{Photon portal \label{app:photonportal}} 
For the photon portal, we can use the same UV completion as in \eqref{eq:gluon_UVcompletion}, with the sole difference that we take the $Q$ fields to be color singlets with hypercharge 1. Upon integrating out the $Q$ fields, we obtain
\begin{equation}\label{eq:photon_portal_intermediate}
\mathcal{L}_{IR}\supset-\frac{1}{2}m_a^2 a^2 -\frac{ \alpha}{4\pi}\frac{y_Q N_Q}{M_Q} a F\tilde F  - i y_\psi a \bar \psi \gamma_5 \psi +\text{h.c.}
\end{equation}
which maps to
\begin{align}
\mathcal{L}_{IR}&\supset \frac{ \alpha}{8\pi}\frac{1}{f_{\heta}}\heta F\tilde F
\end{align}
with $f_{\heta}$ the effective decay constant of $\heta$, given by
\begin{align}\label{eq:fheta_indent_photon}
\frac{1}{f_{\heta}}= \frac{2y_Qy_\psi N_Q \hLambda F_{\heta}}{m_a^2 M_Q}.
\end{align}
Also here there is a certain degree of model dependence in the limits on the vector-like leptons $Q$, depending on their open decay channels. We take $M_Q\gtrsim 700$ GeV as a  rough bound \cite{Sirunyan:2019ofn}, again assuming $y_Q=y_\psi=3$, $N_Q=1$ and $F_{\heta}\approx\hLambda\approx m_{\heta}$. There are no stringent collider constraints on $a$ itself, since the interaction in \eqref{eq:photon_portal_intermediate} is much too weak to produce substantial numbers of $a$ particles through either associated production or photon fusion. We will take $m_a=\text{max}[10\, \text{GeV},2 m_{\heta}]$, such that there are no direct $\heta \to aa$ transitions. With these assumptions we find
\begin{equation}\label{eq:photoneffectivedecay}
f_{\heta}\gtrsim 4\times 10^4\, \mathrm{GeV}\times \left(\frac{m_a}{10\,\mathrm{GeV}}\right)^2 \times \left(\frac{1\;\mathrm{GeV}}{m_{\heta}}\right)^2
\end{equation}
and
\begin{equation}\label{eq:photonctauestimate}
c\tau \gtrsim 0.1\,\mathrm{cm}\times \left(\frac{m_a}{10\,\mathrm{GeV}}\right)^{4} \times \left(\frac{1\;\mathrm{GeV}}{m_{\heta}}\right)^{7}.
\end{equation}
The configuration of the \verb+ctau_min+ function in the python code is analogous to that for the gluon coupling, with $m_a=\text{max}[10\, \text{GeV},2 m_{\heta}]$ as the default choice for the \verb+m_a+ parameter.

\subsection{Vector portal\label{app:vectorportal}} 

\begin{figure}[t]\centering
\includegraphics[width=0.4\textwidth]{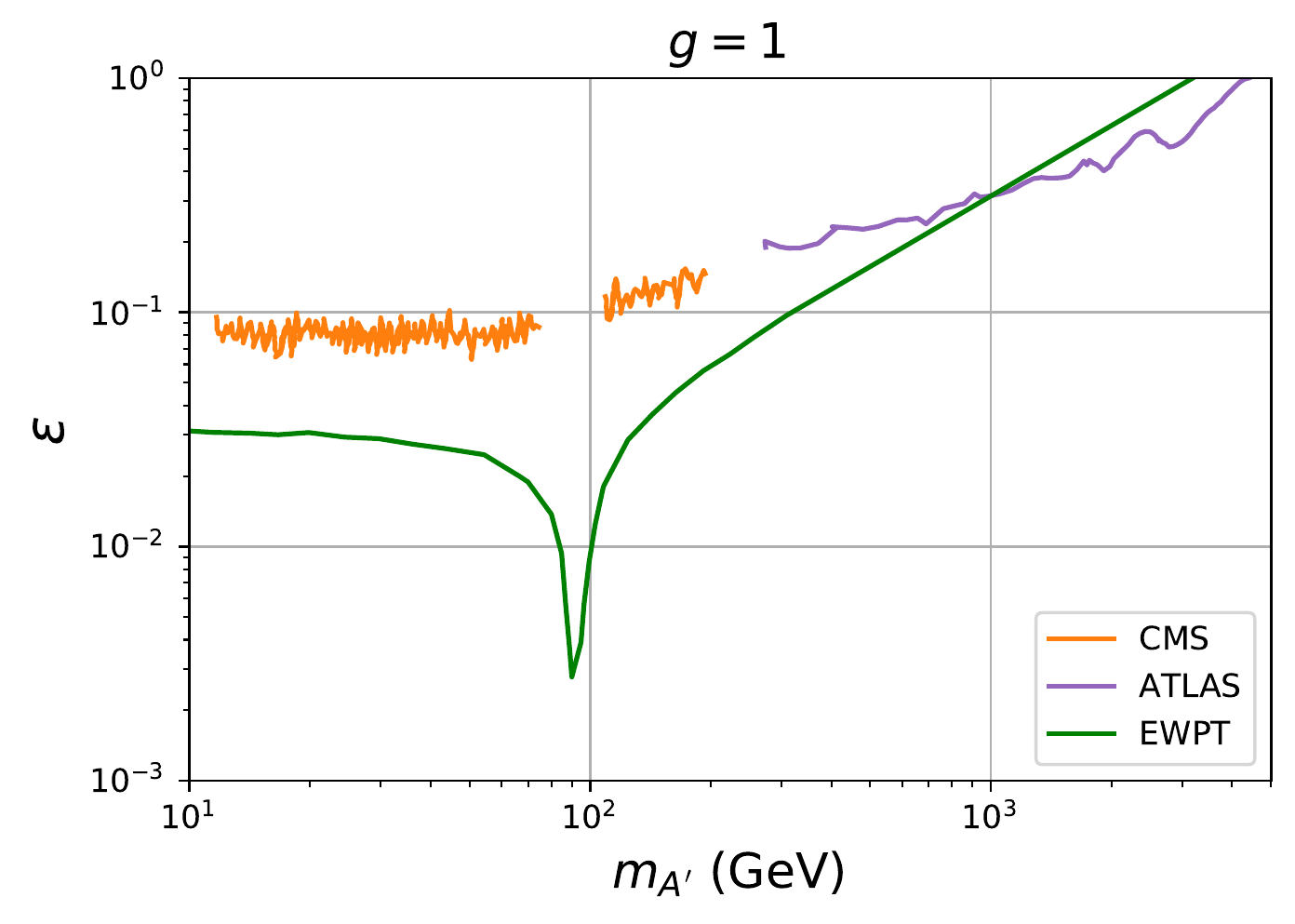}
\caption{Current bounds on a kinetically-mixed dark photon from dilepton searches at CMS \cite{CMS-PAS-EXO-19-018} and ATLAS \cite{Aad:2019fac}, and from electroweak precision tests \cite{Curtin:2014cca}. For $m_{A'}\lesssim 30$ GeV, recent LHCb bounds \cite{Aaij:2019bvg} are slightly stronger than those in \cite{CMS-PAS-EXO-19-018}. Here we take the dark photon to have a partial width to a dark sector specified by coupling to a dark color triplet fermion with dark gauge coupling $g=1$, and show ATLAS limits for a dark photon with a width 10\% of its mass.\label{fig:darkphotonlimits}}
\end{figure}

As explained in Sec.~\ref{sec:vector_portal}, a simple UV completion of the vector portal is a heavy, kinetically-mixed dark photon, which is specified by its mass ($m_{A'}$), its coupling to the dark sector quarks ($g$) and its mixing parameter with the SM photon hypercharge $(\epsilon)$. It is constrained by searches for low and high mass dilepton resonances \cite{Aaij:2019bvg,CMS-PAS-EXO-19-018,Aad:2019fac} and electroweak precision tests (EWPT) \cite{Curtin:2014cca}. When a dark photon can only decay visibly, direct searches are substantially stronger than EWPT constraints; however, in our case the $A'$ decays primarily to the dark sector, which leads to a strong suppression in the $A'\to\ell\ell$ branching ratio. Fig.~\ref{fig:darkphotonlimits} shows our recasted limits, where we assume $g=1$ and a single flavor of dark sector fermions with $N_c=3$ dark colors. For this choice, we find that  EWPT constraints provide the strongest bound in most of the accessible $A'$ mass range. As explained in Sec.~\ref{sec:vector_portal}, we will choose $m_{A'}=500$ GeV and $m_{A'}=750$ GeV for the corresponding signal benchmarks, and saturate the EWPT constraint for these mass points. For our exotic Higgs decay benchmark point, we seek to minimize $\epsilon_{\mathrm{eff}}$ by choosing $m_{A'}=20$ GeV, well outside the kinematical ranges of Belle II and BaBar.

With the above assumptions for the UV completion, the estimated lower bound on the $\homega$ lifetime is
\begin{align}
c\tau &\gtrsim 4\times10^{-2}\,\text{cm} \times \left(\frac{2 \,\text{GeV}}{m_{\homega}}\right)^5\times \left(\frac{m_{A'}}{20\;\mathrm{GeV}}\right)^4\times \left(\frac{10^{-2}}{\epsilon}\right)^2,
\end{align} 
where we assumed $m_{\homega}\approx \hLambda$ and $g\approx 1$.
These bounds are rather loose, especially for the exotic Higgs decay benchmark, due to the very mild constraints on the elementary $A'$ itself. Moreover, the vector model does not require a UV completion with additional charged or colored particles, as was the case for the gluon and photon portals. In our analysis is Sec.~\ref{sec:vector_portal}, we saturate the EWPT bound on $\epsilon$ in  Fig.~\ref{fig:darkphotonlimits} as a function of $m_{A'}$.

 As for the other portals, the lifetime and branching ratios are included in the package, through the functions \verb+ctau(m,eps)+ and \verb+branching_ratios[channel](m)+. Here \verb+m+ and \verb+eps+ are the vector mass $(m_{\homega})$ and effective mixing $\epsilon_{\mathrm{eff}}$. The lower bound on the lifetime can be accessed by \verb+ctau_min(m)+, with optional flags \verb+xi_Lambda+, \verb+m_a+ and \verb+eps+. Here \verb+xi_Lambda+ and \verb+m_a+ respectively fix $\xi_{\Lambda} = \hLambda/m_{\homega}$ and $m_{A'}$, with default values  \verb+xi_Lambda=1+ and  \verb+m_a=20.0+. Here the \verb+eps+ flag indicates the value assumed for $\epsilon$ in \eqref{eq:epseffective}. It default value is set  \verb+eps=-1+, which indicates that the EWPT bound in Fig.~\ref{fig:darkphotonlimits} is used.

\subsection{Higgs portal} 
\label{app:higgsportal}

As for the other portals, in order to write down a concrete UV completion, we must specify the nature of the constituents of the $\heta$ and $\homega$ meson. Our primary goal is to find the UV completion that minimizes the lifetime of the $\heta$.  
For this reason, we here consider a UV completion with \emph{scalar} ``quarks'' $\varphi (\bar\varphi)$ in the fundamental representation of the dark confining gauge group. The $\heta$ can then be identified with the $\varphi\bar \varphi$ bound state and is automatically a scalar, capable of mixing with the SM Higgs without needing to invoke parity violation.  In the UV theory this requires the interactions
\begin{equation}
\label{eq:mass-scalar}
\mathcal{L}_{UV}\supset - m_{\varphi}^2  \varphi \bar \varphi  -\lambda  \varphi \bar \varphi H^\dagger H,
\end{equation}
 with the Higgs portal operator controlled by the dimensionless coupling $\lambda$.\footnote{ Again, we consider scalar constituents in order to estimate the minimum feasible meson lifetime. In the case of fermionic constituents, the UV Higgs portal operator is dimension 5, and meson decay rates are accordingly suppressed by an additional factor of $ (\hLambda/ M)^2$, where $M$ is the scale of the Higgs portal operator.  For glueballs, the UV Higgs portal operator is dimension 6.} However, the Higgs portal operator contributes to the mass of $\varphi$ after electroweak symmetry breaking, so if we refrain from mandating a strong, \emph{tree-level} cancellation between the two terms in (\ref{eq:mass-scalar}),
 we must demand that $\frac{1}{2}\lambda v^2 \lesssim m_\varphi^2 \lesssim \hLambda^2$ with $v=246$ GeV the Higgs vacuum expectation value. In the second equality we demanded that the mass of the dark quarks does not exceed the confinement scale in the dark sector. Saturating this bound and mapping $\varphi\bar\varphi \to \hLambda \heta$, the relevant operator in the IR theory is 
\begin{equation}
\mathcal{L}_{IR}\supset- \frac{2\hLambda^3}{v^2} H^\dagger H \heta,
\end{equation}
which implies a maximum mixing angle between $\heta$ and the Higgs of 
\begin{equation}\label{eq:maxmixangle}
\sin\theta\lesssim \frac{\hLambda^3}{v m_h^2}\sim 5\times10^{-7} \times \left(\frac{\hLambda}{1\text{ GeV}}\right)^3.
\end{equation}

For the lifetime of $\heta$ as a function of $\sin\theta$, we use the results in \cite{Gershtein:2020mwi}, which in turn use the perturbative calculations in \cite{Spira:1997dg} for $m_{\heta}> 2$ GeV and the dispersive computations in \cite{Winkler:2018qyg} for $m_{\heta}<2$ GeV. In the python package, the lifetime and branching ratios of $\heta$ are encoded in the functions \verb+ctau(m,stheta)+ and \verb+branching_ratios[channel](m)+, with \verb+m+ and \verb+stheta+ respectively given by $m_{\heta}$ and $\sin\theta$. The minimum lifetime given by saturating \eqref{eq:maxmixangle} is encoded in the function \verb+ctau_min(m)+, where $\xi_{\Lambda}\equiv \hLambda/m_{\heta}$ can be specified by the optional flag \verb+xi_Lambda+, for which the default value is set to 1.

\bibliography{dark_showers_modelpaper}

\end{document}